\documentclass[prd,superscriptaddress,amsfonts,amssymb,amsmath,showpacs,twocolumn]{revtex4-2}
\usepackage{bm}
\usepackage{amsfonts}
\usepackage{latexsym}
\usepackage{graphicx}
\usepackage{amsmath}
\usepackage{palatino}
\usepackage{mathpazo}
\usepackage{textcomp}
\usepackage{float}
\usepackage{booktabs}
\usepackage{dcolumn}
\usepackage{booktabs}
\usepackage{multirow}
\usepackage{hyperref}
\hypersetup{colorlinks,citecolor=blue}
\usepackage{amsmath}
\usepackage{xcolor}
\usepackage{enumitem}
\usepackage{epsfig}
\usepackage{caption}
\usepackage{subcaption}
\usepackage{commath}
\hypersetup{colorlinks,citecolor=blue}
\hypersetup{colorlinks=trueg,linkcolor=magenta,filecolor=magenta,    urlcolor=blue}

\def\be{\begin{equation}}
\def\ee{\end{equation}}
\def\bea{\begin{eqnarray}}
\def\eea{\end{eqnarray}}

\begin{document}

\title{\bf Strong gravitational lensing by black hole in
$F(R)$-Euler-Heisenberg Gravity’s Rainbow}
\author{Sayan Naskar}
\email{sayannaskar010195@gmail.com}

\author{Niyaz Uddin Molla}
\email{niyazuddin182@gmail.com}

\author{Ujjal Debnath} 
\email{ujjaldebnath@gmail.com} 
\affiliation{ Department of Mathematics, Indian Institute of Engineering Science and
Technology, Shibpur, Howrah-711 103, India }

\begin{abstract}

We investigated gravitational lensing in the strong-field regime for BH in $F(R)$-Euler-Heisenberg Gravity’s Rainbow. This BH spacetime is characterized by the Euler-Heisenberg (EH) parameter $\lambda$, $F(R)$ parameters ($f_{R_0}$, $R_0$), Rainbow parameters ($f_{\epsilon}$, $g_{\epsilon}$), the BH charge $Q$, and mass $M$. We numerically compute the strong deflection angle and its coefficients, then explore astrophysical implications for supermassive BHs in different galaxies.  
Our findings show that increasing $\lambda$ enhances key lensing parameters such as the photon sphere radius $r_{ph}$, critical impact parameter $u_{ph}$, angular position $\theta_{\infty}$, relative magnification $r_{mag}$, Einstein ring radius $\theta^E_{1}$, and time delay $\Delta T_{2,1}$ for a fixed $Q$. Conversely, increasing $Q$  decreases these parameters, keeping the other parameters fixed. Additionally, increasing $\lambda$ reduces the deflection angle $\alpha_D(u)$ and angular separation $S$, whereas increasing $Q$ leads to their increase.  
Our study reveals that BHs in this modified gravity framework act as gravitational lenses, exhibiting deflection angles surpassing those of Reissner-Nordström (RN) and Schwarzschild BHs under specific conditions. Setting $\lambda=0$ retrieves the RN solution, while $R_0=Q=0$ and $f_\epsilon=g_\epsilon=1$ corresponds to Schwarzschild. These findings highlight the distinct topological properties of modified charged BHs and suggest their potential as astrophysical candidates, offering a novel perspective on their observational signatures. The sensitivity of the results has been analyzed to assess how robust the lensing predictions are with respect to changes in the functional forms of the Rainbow functions. Strong gravitational lensing and its key observables have been investigated under the combined modifications of $F(R)$ gravity, Euler–Heisenberg (EH) electrodynamics, and Rainbow Gravity, which together lead to new qualitative effects not present in the individual models.

\end{abstract}



\maketitle
\section{Introduction}
A BH is a region of space with strong gravity that cannot escape, even light or electromagnetic waves\cite{wald2010general}. Einstein's theory of general relativity suggests that a compact mass can create a BH by twisting spacetime\cite{wald1999gravitational,smith2020first}. The point at which there is no way out is known as the event horizon. Although a BH has no locally apparent qualities in accordance with general relativity, it significantly affects the conditions and destiny of an object traveling through it\cite{hamilton2020journey}. As a BH does not reflect any light, it functions similarly to an ideal black body in many respects \cite{schutz2003gravity,davies1978thermodynamics}.

In order to understand the mysterious objects that significantly influence spacetime, BH observations are essential to gravitational and cosmic physics. The LIGO and Virgo Collaboration made history by detecting GW150914, the first gravitational wave event from a binary BH merger\cite{abbott2016properties}. The Event Horizon Telescope (EHT) provided significant proof of the existence of BHs in our universe by capturing an image of the BH at centre of the galaxy M87* \cite{collaboration2019first,akiyama2019event}. These observations open a prospective doorway for more research into the characteristics of BHs.
\par Finding a way to integrate quantum theory and gravitation is one of the biggest problems facing theoretical physics today. Although numerous attempts at unification have been made and are still being considered \cite{carroll2001noncommutative}, none of them are consistent enough to provide a thorough explanation of the idea. The standard energy-momentum dispersion ratio at the ultraviolet limit has been observed as an alternative to Lorentz symmetry violation in various theories, including Horava-Lifshitz \cite{hovrava2009quantum}, loop quantum gravity \cite{amelino2013quantum}, discreteness spacetime \cite{hooft1996quantization}, and doubly special relativity \cite{magueijo2003generalized}. According to doubly special relativity, the dispersion relation can be expressed as follows:
\begin{equation}
 E^2f_{\epsilon}(\epsilon)^2-p^2g_{\epsilon}(\epsilon)^2 = m^2
\end{equation}
Where $f_{\epsilon}(\epsilon)$, $g_{\epsilon}(\epsilon)$ are functions of $\epsilon=E/E_P$, where $E_P$ is the Planck energy and $E$ is the particle energy used to examine the spacetime.The Rainbow functions $f_{\epsilon}(\epsilon)$ and $g_{\epsilon}(\epsilon)$ were created using phenomenology as an inspiration\cite{dehghani2018thermodynamics}. The usual energy dispersion relation was restored by the infrared limit
\begin{equation}
    \lim_{\epsilon\to0}f_{\epsilon}(\epsilon)=\lim_{\epsilon\to0}g_{\epsilon}(\epsilon) = 1.
\end{equation}
It is possible to express a family of energy-dependent metrics (\ref{a}) using a matching set of orthonormal frame fields, each of which is parameterized by a single variable.\cite{magueijo2004gravity}
\begin{equation}\label{a}
    g_{\epsilon}(\epsilon)=\eta^{ab}e_a(\epsilon)e_b(\epsilon)
\end{equation}
Where the frame field's energy dependency is provided by
\begin{equation}
    e_0(\epsilon)=\frac{\tilde{e_0}}{f_{\epsilon}(\epsilon)}
\end{equation}
\begin{equation}
    e_i(\epsilon)=\frac{\tilde{e_i}}{g_{\epsilon}(\epsilon)}
\end{equation}
Energy-independent frame fields are represented by quantities with a tilde. Standard general relativity is recovered in the limit $\epsilon\to0$. A one-parameter set of equations takes the role of Einstein's field equations.
\begin{equation}
    H_{\mu\nu}(\epsilon)=8\pi H(\epsilon)T_{\mu\nu}(\epsilon)
\end{equation}
where $H(\epsilon)$ is a Newton constant that depends on energy. Newton's gravitational constant $H(\epsilon)$, which depends on energy, reduces to the standard Newtonian constant $H=H(0)$ as $\epsilon\to0$. Interestingly, it has been used to provide solutions to the Big Bang singularity by allowing for a non-singular cosmological evolution and avoiding infinite curvature at very high energy, in the early universe. Similarly, Rainbow Gravity has been applied to BH physics\cite{aounallah2022five,dehghani2021quantum}, where it offers a potential mechanism for removing the central singularity predicted by classical general relativity. Although it is still a phenomenological model rather than a complete theory of quantum gravity, Rainbow Gravity captures important characteristics that are anticipated from candidates for quantum gravity, like the presence of a minimum length scale and changes to spacetime at high energies.

 Furthermore, a particle cannot reach energies greater than Planck's energy, which is a constant along with the speed of light. The invariance of the theory is maintained by nonlinear Lorentz transformations, whereas it is broken by linear ones\cite{garattini2015gravity}. Magueijo and Smolin\cite{magueijo2004gravity} generalized doubly special relativity to Rainbow Gravity, where spacetime is represented by parameters in a parameterized metric, making the geometry dependent on the particle's energy being tested. Consequently, a Rainbow of metrics is produced. Since then, some scholars have taken into account the Magueijo along with Smolin theory in their research on generalized theories of gravity\cite{garattini2013distorting,hendi2016f} and BH's research\cite{ling2007thermodynamics,galan2006entropy,liu2007hawking,leiva2009geodesic,farag2014black,ali2015absence,ali2015absence,gim2015black,hendi2016charged,gim2018black,garattini2017gravity,hendi2018ads,panah2018effects}. As a result, the metric is Rainbow-shaped and depends on the particle's energy of the spacetime geometry.

\par A wide range of phenomena can be explained by the F(R) theory, which has generated the most interest among the different modified gravity theories. For example, the F(R) theory of gravity explains the presence of dark matter \cite{abdalla2020dark,nojiri2007introduction} and the universe's fast expansion \cite{faraoni2011beyond,nojiri2011unified,capozziello2011extended,bamba2012dark,joyce2015beyond,nojiri2017modified}. The F (R) hypothesis can also be used to predict huge compact objects \cite{cooney2010neutron,cheoun2013neutron,astashenok2020supermassive,astashenok2021causal,sarmah2022stability,kalita2021gravitational}, cosmic acceleration, or early universe inflation \cite{capozziello2002curvature,aditya2018locally,hu2007models}. Another accomplishment in F(R) gravity is the comprehensive characterization of the universe's evolutionary epochs while preserving conformity to post-Newtonian and Newtonian approximations \cite{capozziello2007newtonian}.In F(R) gravity's context, the theory proposes that F(R) represents a generalized function dependent on the scalar curvature $R_0$. Using the F(R)-Euler-Heisenberg Lagrangian to study and highlight the strong gravitational lensing is crucial because the Rainbow of F(R) gravity may be linked with the Euler-Heisenberg(EH) component, matter of the source, to represent a range of cosmic and astrophysical events. Using Dirac's electron-positron theory, a nonlinear electrodynamics Lagrangian was developed by Heisenberg and Euler \cite{heisenberg2006consequences}.
\par Significant progress has been made in the study of strong gravitational lensing near BHs since the initial exploration of photon trajectories around Schwarzschild BH by Darwin\cite{Darwin:1959md}. In a strong field regime, relativistic images were formalized by Virbhadra and Ellis\cite{Virbhadra:1999nm} while Bozza\cite{Bozza:2002zj} created the strong deflection limit (SDL) for precise light deflection near a BH. These pioneering methods have since been extended to different BH spacetimes, such as those encircled by matter and those in alternative gravity theories \cite{Perlick:2004tq, Bozza:2010xqn, Bambi:2019tjh}.
The lens equation could be solved analytically with the help of Frittelli, Killing, and Newman \cite{Frittelli:1999yf}, and Bozza's SDL framework \cite{Bozza:2001xd, Bozza:2002zj, Bozza:2002af} further developed these techniques. Rotating BH \cite{Bozza:2002af} and Reissner-Nordström BH \cite{Eiroa:2002mk} are two examples of spacetimes to which this technique has been used. Research on alternative BH solutions, including Reissner-Nordström BH, braneworld scenarios, and models within modified gravity frameworks, has also seen substantial advancements due to the application of the strong deflection limit (SDL).\cite{PhysRevD.66.103001, PhysRevD.69.063004, Whisker:2004gq, Eiroa:2012fb, Bhadra:2003zs, Kumar:2022fqo, Kumar:2021cyl, Islam:2021dyk}. Gravitational lensing phenomena in various BH spacetimes, including those described by higher-curvature gravity theories \cite{Kumar:2020sag, Islam:2020xmy, Narzilloev:2021jtg, Islam:2022ybr, Kumar:2020sag} and modifications of Schwarzschild geometries \cite{Eiroa:2010wm, Ovgun:2019wej, Panpanich:2019mll, Bronnikov:2018nub, Molla:2024lpt} are still being investigated in recent work. Recent research has started to investigate how Rainbow functions may alter BH's gravitational lensing characteristics. In recent years, several studies have been conducted on charged BHs, Euler–Heisenberg (EH) electrodynamics, Rainbow Gravity, and various modified gravity theories, each considered individually. These investigations have explored a wide range of physical phenomena, including BH and wormhole shadows, gravitational lensing, accretion disks, quasinormal modes, greybody factors, etc.\cite{Channuie:2025xlw,Okyay:2021nnh,Kuang:2022xjp,Lambiase:2024lvo,Ditta:2024ccz,Verbin:2024ewl,Gogoi:2024epx,Pulice:2023dqw,Ovgun:2025stp,AraujoFilho:2024mvz}. However, in the present work, we aim to study the gravitational lensing effects arising from the combined influence of charge, EH electrodynamics, Rainbow Gravity, and $F(R)$-type modified gravity within a unified framework.
\par This study introduces the Rainbow profile of $F(R)$-EH gravity around a BH to enhance our comprehension of strong gravitational lensing. Comparing the traditional lensing situation surrounding a BH in a vacuum with the existence of $F(R)$-EH Gravity’s Rainbow, we analyze the changes in important lensing observables, including the deflection angle, image magnifications, and time delays. The findings have significant implications for evaluating GR and alternative gravity theories, understanding BH environments, and using astrophysical evidence to study $F(R)$ and Rainbow functions characteristics.
Our goal is to examine gravitational lensing caused by an $F(R)$-EH Rainbow BH in strong field approximations. We compare the unique features of this BH with those of an RN BH ($f_{R_{0}}=0, R_{0}=4\Lambda,g_{\epsilon}=1,f_{\epsilon}=1$ and $\lambda=0$) and the Schwarzschild BH as well ($R_{0}=Q=0$ and $f_{\epsilon}=g_{\epsilon}=1$). Our goal is to detect distinct characteristics in strong lensing situations while elucidating the complex interactions between the EH parameter $\lambda$, electric charge Q, and gravitational field. We specifically examine how the charged parameter Q and the EH parameter $\lambda$ affect the deflection angle, time delay between relativistic images, and strong lensing observables when the other parameters are fixed. For the strong field regime, we use Bozza \cite{bozza2010gravitational} methods for gravitational lensing. 

The structure of this paper is as follows: In Section \ref{sec2}, a $F(R)$-Euler Heisenberg Rainbow BH spacetime is reviewed. In section \ref{sec3}, we derive null geodesics and the effective potential of $F(R)$-EH gravity's Rainbow BH. Strong gravitational lensing by an $F(R)$-EH gravity's Rainbow BH is examined in Section \ref{sec4}. Strong lensing observables, such as angular image position, separation, relative magnification, Einstein's ring, and the time delays of relativistic images, are also examined in this section. Astrophysical bound of EH parameter using EHT observation data for M87* and SgrA*, are obtained in section~\ref{sec5} . Furthermore,observational signatures of gravitational lensing in different rainbow function frameworks have been studied in  section ~\ref{sec6}. Finally, we include the concluding remarks of the obtained results in section \ref{sec7}.

\section{$F(R)$-Euler-Heisenberg (EH) Gravity’s Rainbow:}\label{sec2}
In the context of the $F(R)$-EH gravity's Rainbow, considering the energy-dependent, spherically symmetric, static four-dimensional spacetime:
 \begin{equation}
ds^2 = -G(r)f_{\epsilon}^{-2} dt^2 + g_\epsilon^{-2}\Big(G(r)^{-1} dr^2 + r^2 ( d\theta^2 + \sin^2 \theta d\phi^2)\Big)
\label{metric} 
\end{equation}
  Metric function $ G(r) $ is provided by\cite{sekhmani2024thermodynamic}:
\begin{equation} 
 G(r) = 1-\frac{M}{r}-\frac{R_{0}r^2}{12g_{\epsilon}^2}+\frac{f_{\epsilon}^2}{1+f_{R_{0}}}\Big(\frac{Q^2}{r^2}-\frac{\lambda Q^4}{20 r^6}\Big)
\label{fin_metric_funct}
\end{equation}
Integrating constant M is connected to BH's geometric mass,$f_{\epsilon}$ and $g_{\epsilon}$ are Rainbow functions, and Q,$\lambda$ are respectively charge and EH  parameter of the BH.Also the $F(R)$ parameters  $R_{0}$ is called constant scalar curvature\cite{cognola2005one} and $f_{R_{0}}=f_{R_{|R=R_{0}}}$,is a constraint parameter. If we consider the particular partial space,$f_{R_{0}}=0$,$R_{0}=4\Lambda$ and $f_{\epsilon}=g_{\epsilon}=1$ and then the metric function $G(r)$ reduces to-
\begin{equation}
    G(r)=1-\frac{M}{r}-\frac{\Lambda r^2}{3}+\frac{Q^2}{r^2}-\frac{\lambda Q^4}{20 r^6}
\end{equation}
This is the well known Euler-Heisenberg -(A)dS BH, and the Schwarzschild BH is represented if $R_{0}=Q=0$ and $f_{\epsilon}=g_{\epsilon}=1$.

Investigating the set of corresponding real roots is made possible by looking at the metric function. However, by analyzing the appropriate set of these actual roots, it is possible to learn more about the horizons, both the inner and outer horizons, as well as the cosmological horizon. In order to continue with this analysis, a numerical method is also required \cite{sekhmani2024thermodynamic}.
\par It is possible to specify the horizon radii spectrum by solving the equation $G(r = r_h) = 0$. For the change of metric function as a function of the space-time variable r, the appropriate horizon radii can be expressed as either $R_0<0$ (or AdS space if $R_0 = 4\Lambda$) or $R_0 > 0$ (or dS space if $R_0 = 4\Lambda$).The horizon structure provides a maximum of four horizon radii (including the cosmological horizon) when $R_0 > 0$.But, the $ R_0 < 0$ scenario can only include the cosmic horizon and the other three potential horizon radii\cite{sekhmani2024thermodynamic}. The BH horizon's structure in the $F(R)$-EH gravity's Rainbow is largely determined by the sign of the $F(R)$ parameter $R_0$.
    \begin{figure*}
	\begin{centering}
		\begin{tabular}{p{10cm} p{10cm}}
		\includegraphics[scale=1]{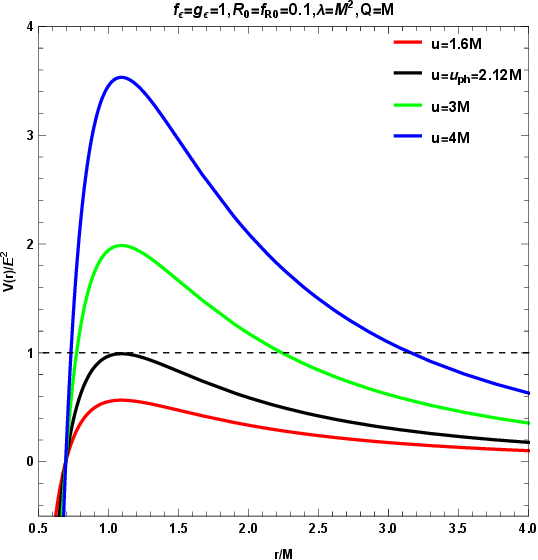}
			\end{tabular}
	\end{centering}
	\caption{  Particular values of the constant scalar curvature $R_{0}=0.1$, the constant parameter $f_{R_{0}}=0.1$, EH parameter $\lambda=M^2$ and charge $Q=M$  are used to determine the effective potential's functional form, $V_{eff}(r)$, as a function of the radial distance $r$. The plot shows the behavior of the effective potential for varying u. Blue, green, black, and red, respectively, represent the trajectories that correspond to values of $u$ greater than $u_{ph}$, equal to $u_{ph}$, and less than $u_{ph}$ }
    .\label{fig1}	
\end{figure*}
\section{Null geodesics of $F(R)$-EH Rainbow BH}\label{sec3}
The metric function can be written as:
\begin{equation}
ds^2=-  f(r)  dt^2 +g(r) dr^2+
h(r)(d\theta^2+sin^2\theta d\phi^2)
\label{1}
\end{equation}
 The functions $f(r)$, $g(r)$, and $h(r)$ are compared to Eq.(\ref{metric}), resulting in the following equations.

 \begin{equation}
f(r) = \frac{G(r)}{f_{\epsilon}^2},
\end{equation}
\begin{equation}
g(r) = \frac{1}{G(r)g_{\epsilon}^2},
\end{equation}

\begin{equation}
h(r)=\frac{r^{2}}{g_{\epsilon}^2}.
\end{equation}

 Regarding the situation of F(R)-EH gravity's Rainbow BH, the motion of photons is governed by the Lagrangian $\mathcal{L} = -\frac{1}{2}g_{\mu \nu}\dot{x}^{\mu}\dot{x}^{\nu}$.
 Assuming that the photon’s trajectory exists only on the equatorial plane, we set $\theta = \frac{\pi}{2}$. Given the F(R)-EH Rainbow spacetime metric, as represented by equation (\ref{1}), the corresponding Lagrangian equation describing photon motion within this spacetime can be expressed as follows:

  \begin{equation}\label{2}
  \begin{split}
 & \mathcal{L}=-\frac{1}{2}g_{\mu \nu}\dot{x}^{\mu}\dot{x}^{\nu}\\
& =\frac{1}{2}\Big(f(r)dt^2-g(r) dr^2 -h(r) (d\theta^2 + \sin^2\theta d\phi^2)\Big)=\varepsilon. \\
 \end{split}
  \end{equation}

The photon's four-velocity is denoted by $\dot{x}^{\mu}$, where the dot represents differentiation with respect to the affine parameter $\tau$.$\varepsilon$ is a parameter with values of $-1$, $0$, and $1$, representing spacelike, null, and timelike geodesics, respectively. A photon orbiting the $F(R)$-EH Rainbow BH has a null geodesic of $\varepsilon = 0$.

The equations regulating the null geodesics can be expressed as follows using Eq. (\ref{2}):

   \begin{equation}\label{3}
   \dot{t}=\frac{dt}{d\tau}=\frac{E}{f(r)},
       \end{equation}

   \begin{equation}\label{4}
   \dot{\phi}= \frac{d\phi}{d\tau}=\frac{L}{r^2},
   \end{equation}
and 
   \begin{equation}\label{5}
   \dot{r}=\frac{dr}{d\tau} =\pm\sqrt{\frac{1}{g(r)}\biggr(\frac{E^2 }{f(r)}-\frac{L^2}{r^2}\biggr)}.
   \end{equation}

The outgoing photon motion is shown by the symbol '+', and the symbol '-' indicates the incoming photon motion. The energy and angular momentum of a photon are denoted by the variables $E$ and $L$, respectively. Eq. (\ref{5}) can be written as:

\begin{equation}
    \dot{r}^2+V_{eff}(r)=0
\end{equation}
where the photon's effective potential is provided by

\begin{equation}
    V_{eff}(r)=\frac{1}{g(r)}\biggr(\frac{L^2}{r^2}-\frac{E^2 }{f(r)}\biggr)
\end{equation}
A photon's critical ring orbit is defined by the effective potential function $V_{eff}(r)$, which satisfies the certain conditions, $\frac{dV_{eff}(r)}{dr}=0$,$\frac{d^2V_{eff}(r)}{dr^2}>0$ for stable circular orbit and $\frac{d^2V_{eff}(r)}{dr^2}<0$ for unstable circular orbit.
It should be noted that the F(R)-EH Rainbow BH satisfies the criterion $\frac{d^2V_{eff}(r)}{dr^2}<0$ for an unstable circular orbit. A photon sphere with radius $r_{ph}$ is therefore created when photon rays arrive from infinity and move in a circular orbit (unstable) around the F(R)-EH Rainbow BH. 
The radius of the photon sphere $r_{ph}$ is determined by calculating the greatest real root of the following equation \cite{Virbhadra:2002ju, Claudel:2000yi}:
\begin{equation}
    \frac{f(r)}{f'(r)}=\frac{h(r)}{h'(r)}
\end{equation}

\begin{figure*}[htbp]
 \captionsetup[subfigure]{labelformat=simple}
    \renewcommand{\thesubfigure}{(\alph{subfigure})}
		\begin{subfigure}{.45\textwidth}
			\caption{}\label{fig2a}
			\includegraphics[height=3in, width=3in]{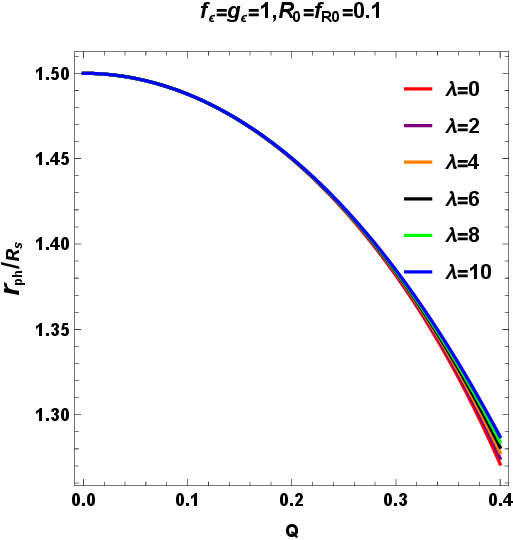}
		\end{subfigure}
            \begin{subfigure}{.4\textwidth}
			\caption{}\label{fig2b}
			\includegraphics[height=3in, width=3in]{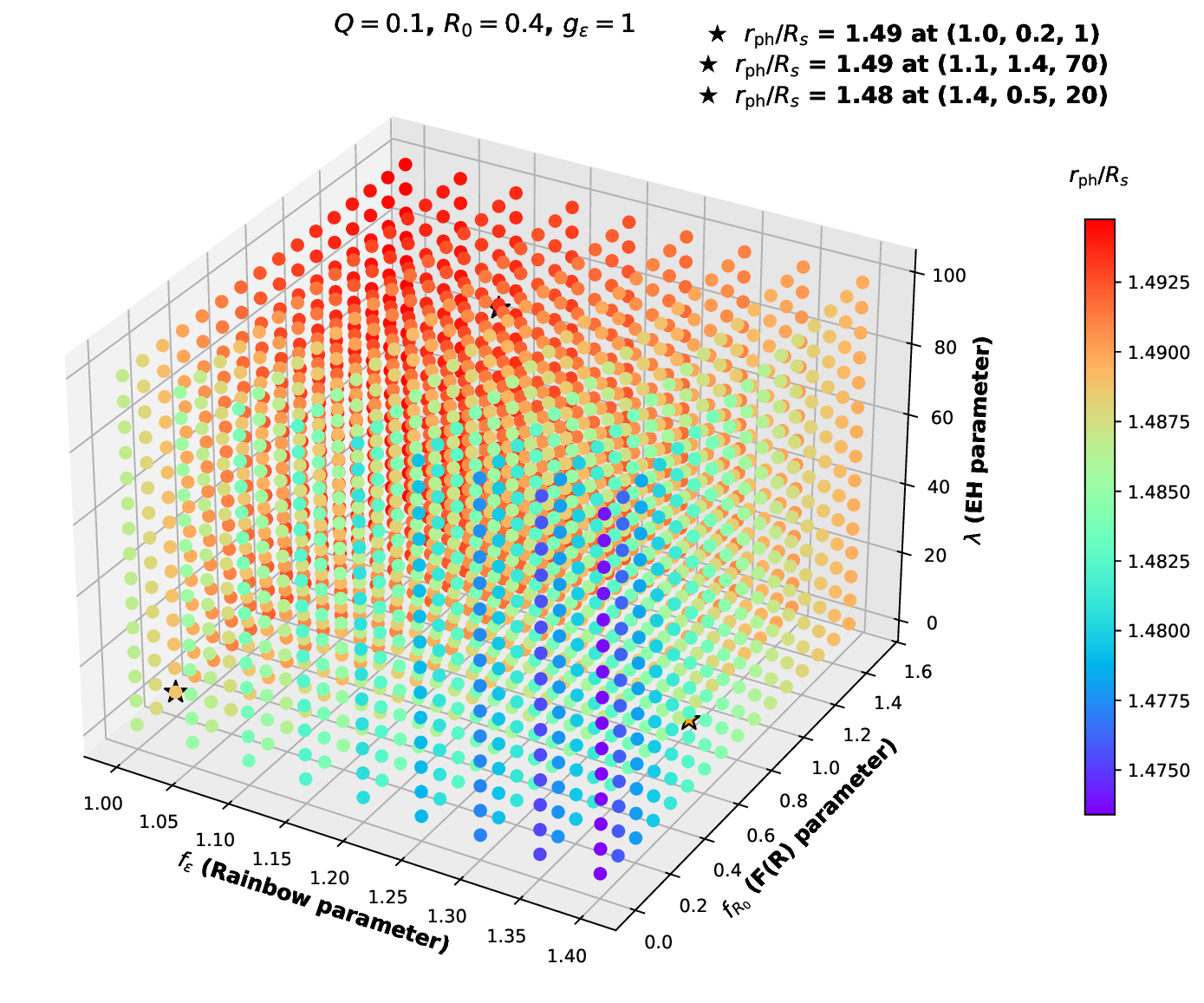}
		\end{subfigure}
  
		\caption{(a) Behavior of the photon sphere $r_{ph}$ with EH parameters $\lambda$ and charge parameter Q, keeping the other parameters fixed. (b) The photon sphere $r_{ph}$ is significantly influenced by the combined modifications from $F(R)$ gravity, Euler-Heisenberg electrodynamics (EH), and Gravity's Rainbow.}
		\label{fig2}
\end{figure*}

The photon sphere radius $r_{\text{ph}}$ is computed in a dimensionless form as $r_{\text{ph}} / R_s$, as shown in Fig.~\ref{fig2}, where $R_s $ defined as $R_s = \frac{GM}{c^2}$ \cite{Virbhadra:1999nm}. Fig.~\ref{fig2a} illustrates the variation of the photon sphere $r_{\text{ph}}$ as a function of two parameters: the EH parameter $\lambda$ and the charge parameter $Q$. In contrast, Fig.~\ref{fig2b} demonstrates the dependence of $r_{\text{ph}}$ on three parameters: the EH parameter $\lambda$, the $F(R)$ gravity parameter $f_{R_0}$, and the Rainbow parameter $f_{\epsilon}$, highlighting how $r_{\text{ph}}$ is influenced by the combined modifications from $F(R)$ gravity, EH electrodynamics, and Rainbow Gravity.
For a fixed value of the EH parameter $\lambda$, while keeping all other parameters constant, Fig.~\ref{fig2a} clearly shows that the photon sphere radius $r_{\text{ph}}$ decreases with increasing charge parameter $Q$. Conversely, for a fixed value of $Q$, the photon sphere radius $r_{\text{ph}}$ increases as $\lambda$ increases. Notably, when $Q = 0 = R_0$ and $f_{\epsilon}=g_\epsilon=1$, the photon sphere radius reduces to $r_{\text{ph}} / R_s = 1.5$, which corresponds to the Schwarzschild BH, consistent with the result in \cite{Bozza:2002zj}.
The behavior depicted in Fig.~\ref{fig2b} motivates the simultaneous consideration of $F(R)$ gravity, Euler–Heisenberg electrodynamics, and Rainbow Gravity, as their combined effects yield nontrivial modifications to the photon sphere structure.

\section{Strong Gravitational lensing  by $F(R)$-EH gravity's Rainbow BH}\label{sec4}
In order to examine the effects of the charged parameter Q, EH parameter $\lambda$, Rainbow functions $f_{\epsilon}$, and $g_{\epsilon}$ on the strong lensing observables, we compare the strong gravitational lensing in an EH Rainbow BH using null geodesics to both the Schwarzschild BH $(R_{0}=0=Q,f_{\epsilon}=g_\epsilon=1)$ and the Reissner–Nordström BH ($f_{R_{0}}=0$,$R_{0}=4\Lambda$,$f_{\epsilon}=g_{\epsilon}=1$ and $\lambda=0$).
The effect of an $F(R)$-EH Rainbow BH on photon ray deflection on the equatorial plane $(\theta=\frac{\pi}{2})$ is investigated here.

\par Using the unit of Schwarzschild radius $\frac{GM}{c^2}$ \cite{Virbhadra:1999nm,drvirbhaa2000schwarzschild}, (where $G=1=c$), we scale the parameters $r$, $t$, $\lambda$, $R_{0}$, and $Q$ in order to recast the metric in Eq. (\ref{1}) and obtain the strong deflection angle for light beams in the BH equatorial plane. Specifically, the transformations are: $t\rightarrow t/M$, $r\rightarrow
r/M$, $q \rightarrow q/M$,$\lambda \rightarrow \lambda/M^2$ and $R_{0} \rightarrow M^2 R_0 $ , respectively
\begin{equation}
ds^2=-  A(r)  dt^2 +B(r) dr^2+
C(r)d\phi^2
\label{act2}
\end{equation}
where 
\begin{equation}
    A(r)=\Big( 1-\frac{1}{r}-\frac{R_{0}r^2}{12g_{\epsilon}^2}+\frac{f_{\epsilon}^2}{1+f_{R_{0}}}\Big(\frac{Q^2}{r^2}-\frac{\lambda Q^4}{20 r^6}\Big)\Big)/f_\epsilon^2
\end{equation}
\begin{equation}
    B(r)=\frac{(A(r))^{-1}}{g_{\epsilon}^2}
\end{equation}
\begin{equation}
    C(r)=\frac{r^2}{g_\epsilon^2}
\end{equation}

\begin{equation}
    u=\frac{L}{E}=\sqrt{\frac{C(r)}{A(r)}}
\end{equation}
One can determine the value of $u_{0}$ when a particle reaches the closest approach distance $r_0$ to the BH, where $\frac{dr}{d\tau} = 0$ \cite{PhysRevD.66.103001}, as:
\begin{equation}
    u_{0}=\sqrt{\frac{C(r_{0})}{A(r_{0})}}
\end{equation}
For the unstable photon orbit, the critical impact parameter $u_{ph}$ is provided by,
\begin{equation}
    u_{ph}=\sqrt{\frac{C(r_{ph})}{A(r_{ph})}}
\end{equation}

\begin{figure*}[htbp]
 \captionsetup[subfigure]{labelformat=simple}
    \renewcommand{\thesubfigure}{(\alph{subfigure})}
		\begin{subfigure}{.45\textwidth}
			\caption{}\label{fig3a}
			\includegraphics[height=3in, width=3in]{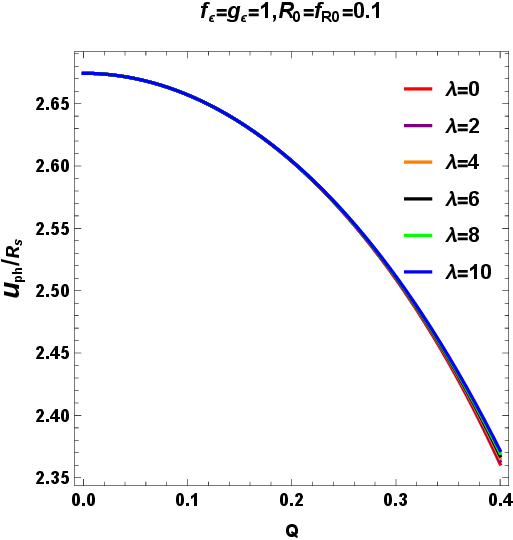}
		\end{subfigure}
            \begin{subfigure}{.4\textwidth}
			\caption{}\label{fig3b}
			\includegraphics[height=3in, width=3in]{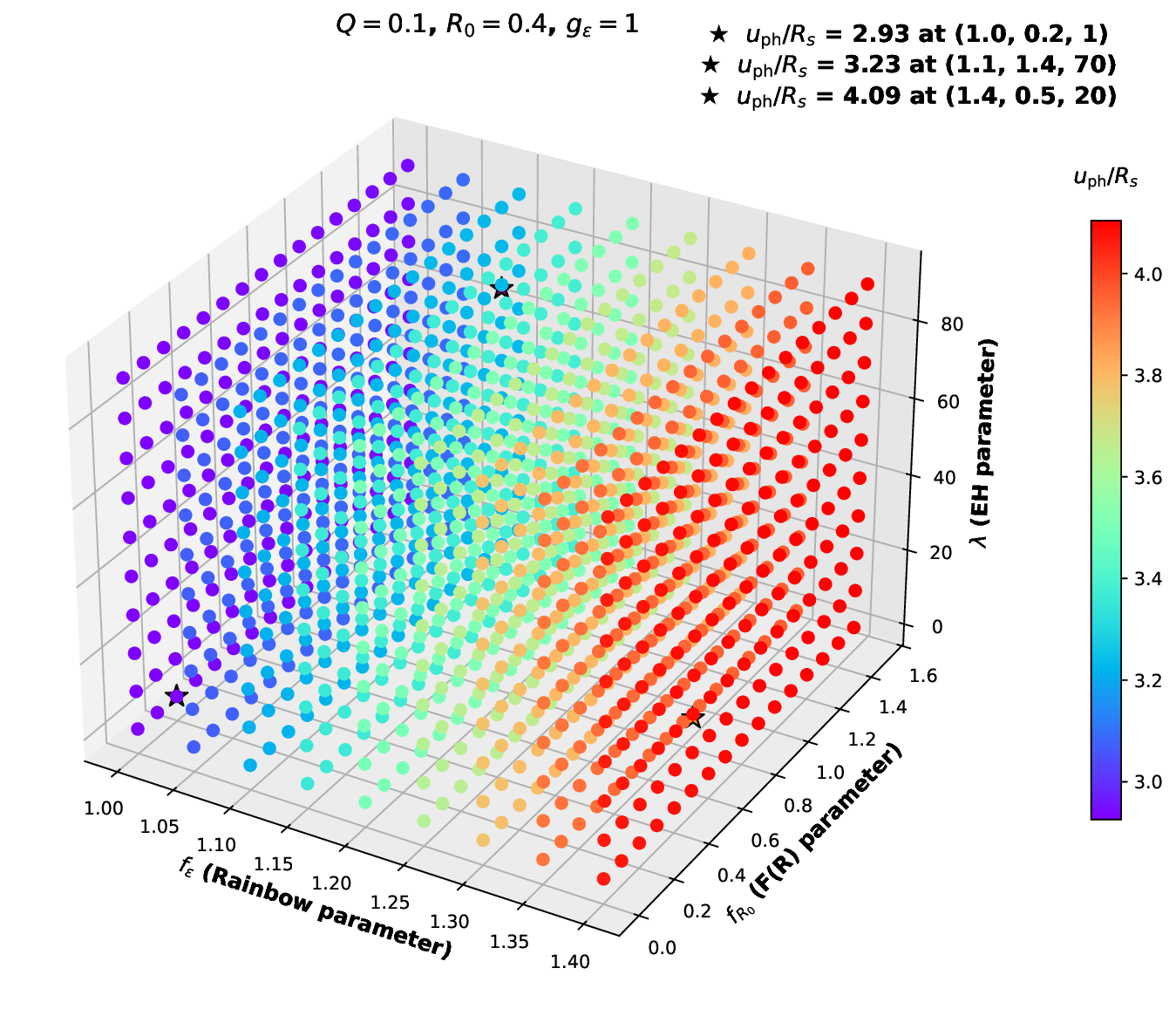}
		\end{subfigure}
  
		\caption{(a) Behavior of the critical impact parameter $u_{ph}$ with EH parameters $\lambda$ and charge parameter Q, keeping the other parameters fixed. (b) The critical impact parameter $u_{ph}$ is significantly influenced by the combined modifications from $F(R)$ gravity, Euler-Heisenberg electrodynamics (EH), and Gravity's Rainbow.}
		\label{fig3}
\end{figure*}

The critical impact parameter $u_{ph}$ is computed in a dimensionless form as $u_{\text{ph}} / R_s$, as shown in Fig.~\ref{fig3} and numerically calculated in Table \ref{t1}. Fig.~\ref{fig3a} illustrates the variation of the critical impact parameter $u_{ph}$ as a function of two parameters: the EH parameter $\lambda$ and the charge parameter $Q$. In contrast, Fig.~\ref{fig3b} demonstrates the dependence of $u_{\text{ph}}$ on three parameters: the EH parameter $\lambda$, the $F(R)$ gravity parameter $f_{R_0}$, and the Rainbow parameter $f_{\epsilon}$, highlighting how $u_{\text{ph}}$ is influenced by the combined modifications from $F(R)$ gravity, EH electrodynamics, and Rainbow Gravity.
For a fixed value of the EH parameter $\lambda$, while keeping all other parameters constant, Fig.~\ref{fig3a} clearly shows that the critical impact parameter $u_{ph}$ decreases with increasing charge parameter $Q$. Conversely, for a fixed value of $Q$, the critical impact parameter $u_{ph}$ increases as $\lambda$ increases. Notably, when $Q = 0 = R_0$ and $f_{\epsilon}=g_\epsilon=1$, the critical impact parameter reduces to $u_{\text{ph}} / R_s = 2.59808$, which corresponds to the Schwarzschild BH, consistent with the result in \cite{Bozza:2002zj}.
The behavior depicted in Fig.~\ref{fig3b} motivates the simultaneous consideration of $F(R)$ gravity, Euler–Heisenberg electrodynamics, and Rainbow Gravity, as their combined effects yield nontrivial modifications to critical impact parameter $u_{ph}$ value.

   For the $F(R)$-EH Rainbow BH spacetime, the following formula can be used to get the strong field limit's deflection angle:
\cite{Virbhadra:1998dy}
\begin{equation}
    \alpha_{D}(r_{0})=I(r_{0})-\pi
\end{equation}
where $I(r_{0})$ is given by:
\begin{equation}\label{de1}
    I(r_{0})=2\int_{r_{0}}^{\infty}\frac{\sqrt{B(r)}}{\sqrt{C(r)}\sqrt{\frac{A(r_{0})C(r)}{A(r)C(r_{0})}-1}}dr
\end{equation}
The relationship between $r_{0}$ and $r_{ph}$ determines the strong deflection angle $\alpha_{D}(r_{0})$, which is increased when $r_{0}\approx r_{ph}.$
In order to avoid  integrating the Eq. \ ref{de1} up to infinity, we establish a new variable z in the manner shown below \cite{tsukamoto2017deflection,tsukamoto2016strong}:

\begin{equation}
    z=1-\frac{r_{0}}{r}
\end{equation}
The strong deflection angle in the scenario when $r_{0}\approx r_{ph}$ is finally can be expressed as

\begin{equation}\label{24}
    \alpha_{D}(u)=-\bar{a} log \Big(\frac{u}{u_{ph}}-1\Big)+\bar{b}+\mathcal{O}\big((u-u_{ph}) log (u-u_{ph})\big)
\end{equation}
 the strong lensing coefficient $\bar{a}$, $\bar{b}$ are given respectively:
\begin{equation}
    \bar{a}=\sqrt{\frac{2A(r_{ph})B(r_{ph})}{A(r_{ph})C''(r_{ph})-A''(r_{ph})C(r_{ph})}}
\end{equation}
and
\begin{equation}
    \bar{b}=-\pi+I_{R}(r_{ph})+\bar{a}log\Big[r_{ph}^2\Big(\frac{C''(r_{ph})}{C(r_{ph})}-\frac{A''(r_{ph})}{A(r_{ph})}\Big)\Big]
\end{equation}
Here,
\begin{equation}
  \begin{split}
     I_{R}(r_{ph})=&2\int_{0}^{1}r_{ph}\Bigg(\Bigg[\sqrt{\frac{B(z)}{C(z)}}\Big(\frac{A(r_{ph})C(z)}{A(z)C(r_{ph})}-1\Big)\frac{1}{(1-z)^2}\Bigg]\\
     & -\frac{\bar{a}}{zr_{ph}}\Bigg)dz
    \end{split}
\end{equation}

\begin{table*}
 \caption{ Strong lensing coefficients are estimated using various values of the charge EH parameter $Q=0,0.1,0.2,0.3,0.4,0.5$ and the parameter $\lambda=0,1,6,12$ for the $F(R)$-EH gravity's Rainbow BH.}\label{t1}
 \begin{tabular}{p{3.2cm} p{3.5cm} p{3.5cm}p{3cm} p{2cm} p{3cm} p{2cm}} 
\hline
\hline
\multicolumn{5}{c}{Strong Lensing Coefficients }\\
$\lambda$ & Q & $\bar{a}$  & $\bar{b} $ &$\ u_{ph}/R_s$\\
\hline
SchwarzSchild BH ($Q=0,R_0=0$) & & 1 & -0.40023 & 2.59808 \\
\hline

1    & 0.1 &  1.00301 & -0.146556  & 2.93424 \\
     & 0.2 &  1.01268 & -0.150425  & 2.8826  \\
     & 0.3 &  1.0313  & -0.157685  & 2.79388 \\
     & 0.4 &  1.06481 & -0.171144  & 2.66273 \\
     & 0.5 &  1.1309  &  -0.203417 & 2.47687 \\
\hline
     & 0.1 &  1.00299 & -0.146537 & 2.93425 \\
6    & 0.2 &  1.01235 & -0.149872 & 2.88277 \\
     & 0.3 &  1.02913 & -0.153964 & 2.79494 \\
     & 0.4 &  1.0544  & -0.152767 & 2.66714 \\
     & 0.5 &  1.08326 & -0.117176 & 2.49349 \\   

\hline
     & 0.1 &  1.00298 & -0.146507 & 2.93426\\
12   & 0.2 &  1.01201 & -0.149303 & 2.88295 \\
     & 0.3 &  1.02691 & -0.150209 & 2.79603 \\
     & 0.4 &  1.04439 & -0.135746 & 2.67159 \\
     & 0.5 &  1.04652 & -0.0598826& 2.50901\\
\hline
\hline
\end{tabular}
\end{table*}

\begin{figure*}[htbp]
 \captionsetup[subfigure]{labelformat=simple}
    \renewcommand{\thesubfigure}{(\alph{subfigure})}
		\begin{subfigure}{.45\textwidth}
			\caption{}\label{fig4a}
			\includegraphics[height=3in, width=3in]{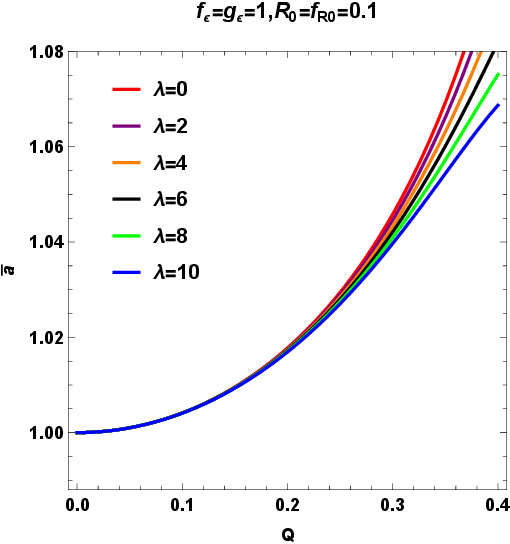}
		\end{subfigure}
            \begin{subfigure}{.4\textwidth}
			\caption{}\label{fig4b}
			\includegraphics[height=3in, width=3in]{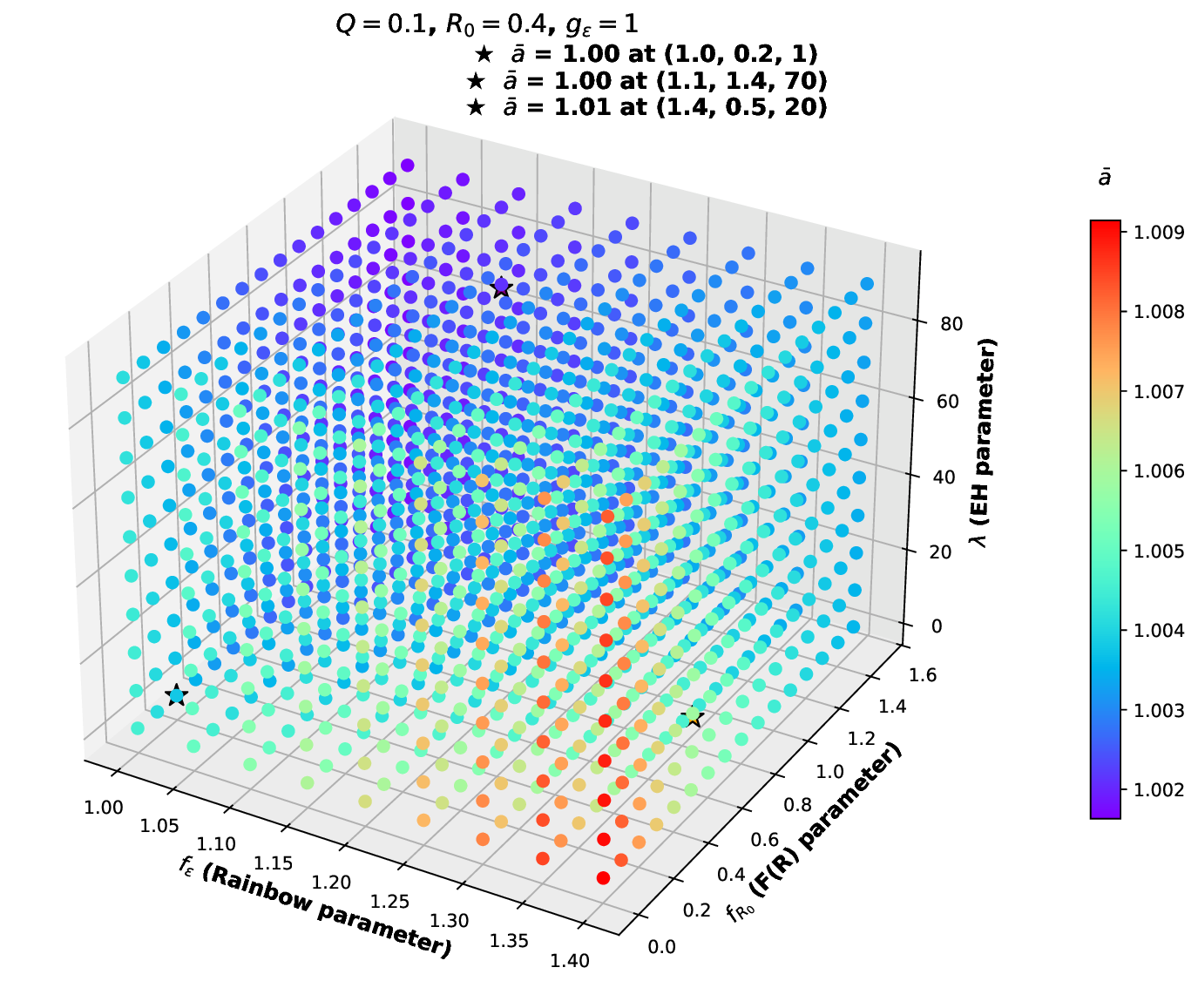}
		\end{subfigure}
  
		\caption{ (a) Behavior of the strong lensing coefficient $\bar{a}$ with EH parameters $\lambda$ and charge parameter Q, keeping the other parameters fixed. (b) The strong lensing coefficient $\bar{a}$ is significantly influenced by the combined modifications from $F(R)$ gravity, Euler-Heisenberg electrodynamics (EH), and Gravity's Rainbow.}
		\label{fig4}
\end{figure*} 

\begin{figure*}[htbp]
 \captionsetup[subfigure]{labelformat=simple}
    \renewcommand{\thesubfigure}{(\alph{subfigure})}
		\begin{subfigure}{.45\textwidth}
			\caption{}\label{fig5a}
			\includegraphics[height=3in, width=3in]{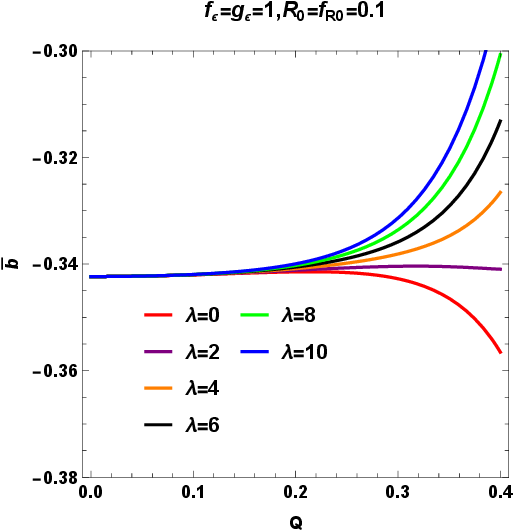}
		\end{subfigure}
            \begin{subfigure}{.4\textwidth}
			\caption{}\label{fig5b}
			\includegraphics[height=3in, width=3in]{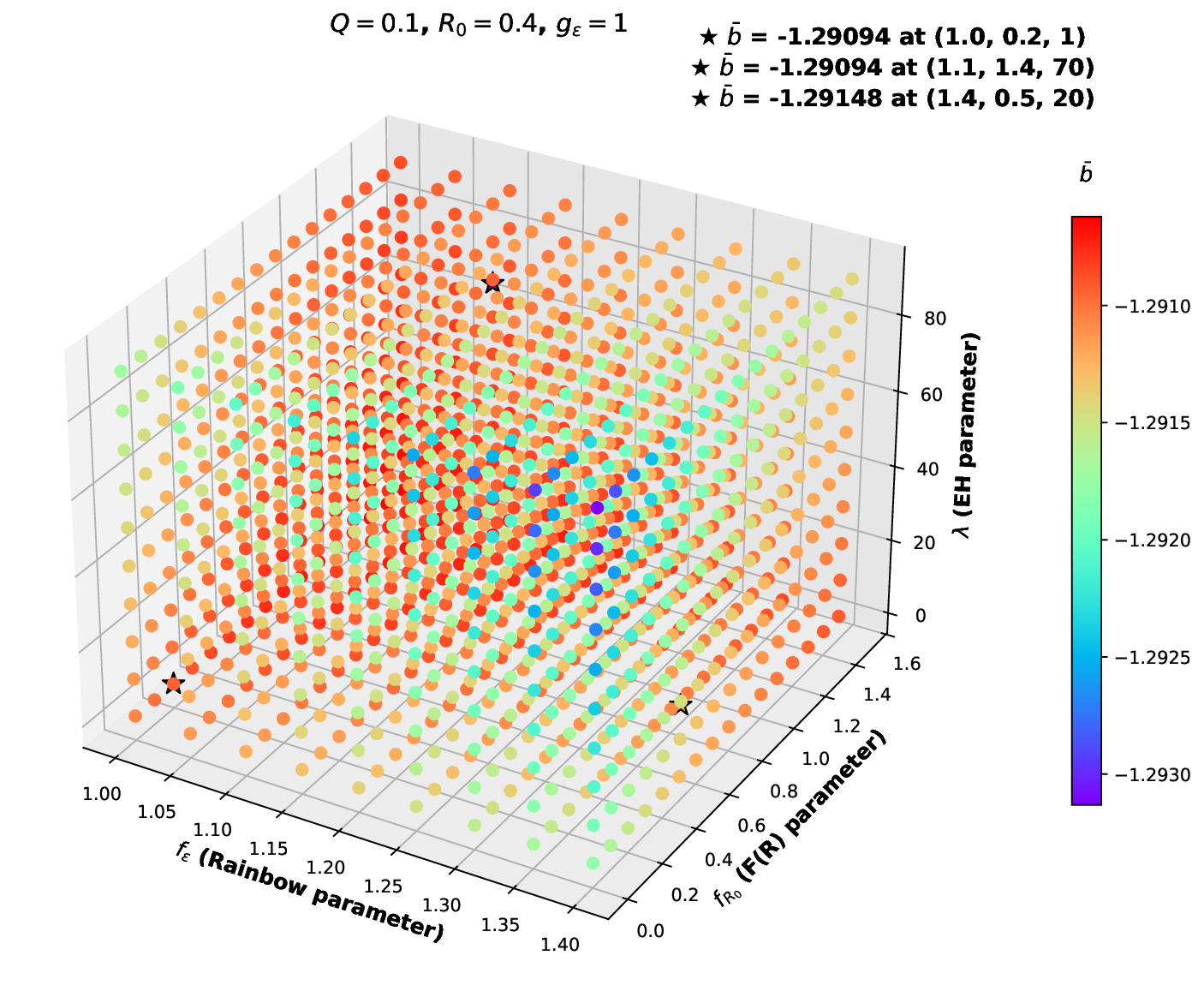}
		\end{subfigure}
  
		\caption{(a) Behavior of the strong lensing coefficient $\bar{b}$ with EH parameters $\lambda$ and charge parameter Q, keeping the other parameters fixed. (b) The strong lensing coefficient $\bar{b}$ is significantly influenced by the combined modifications from $F(R)$ gravity, Euler-Heisenberg electrodynamics (EH), and Gravity's Rainbow.}
		\label{fig5}
\end{figure*}

\begin{figure*}[htbp]
 \captionsetup[subfigure]{labelformat=simple}
    \renewcommand{\thesubfigure}{(\alph{subfigure})}
		\begin{subfigure}{.45\textwidth}
			\caption{}\label{fig6a}
			\includegraphics[height=3in, width=3in]{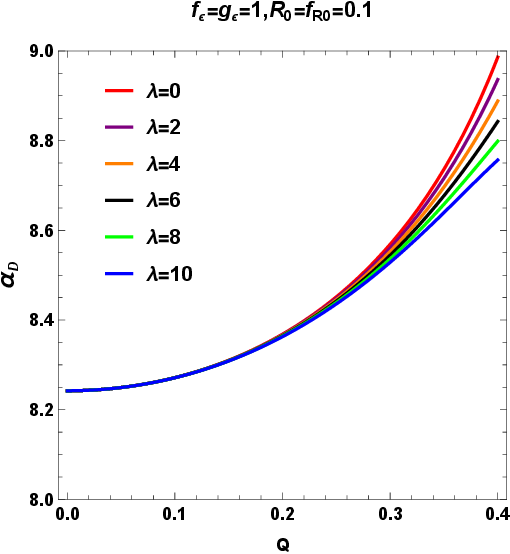}
		\end{subfigure}
            \begin{subfigure}{.4\textwidth}
			\caption{}\label{fig6b}
			\includegraphics[height=3in, width=3in]{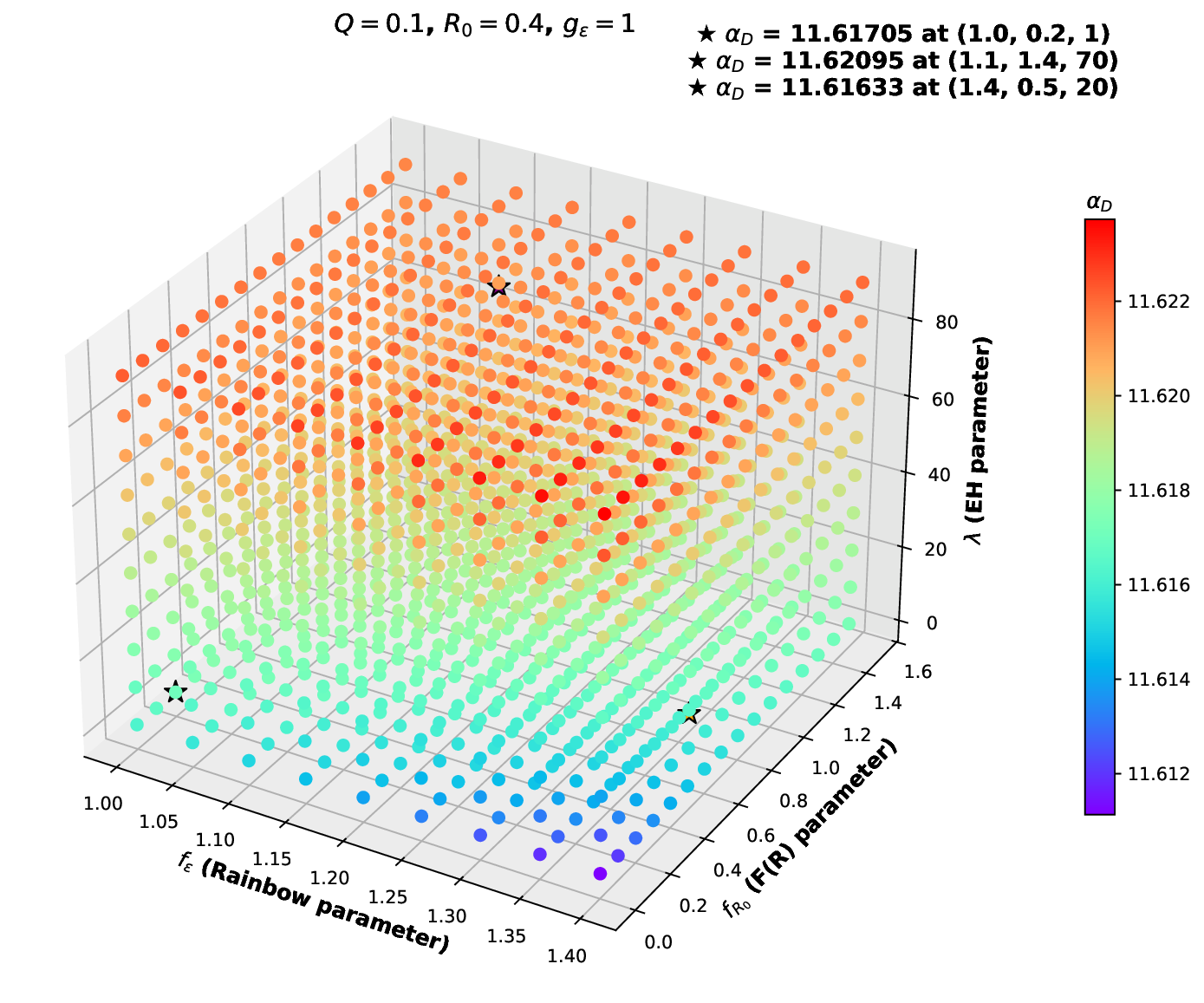}
		\end{subfigure}
  \begin{subfigure}{.45\textwidth}
			\caption{}\label{fig6c}
			\includegraphics[height=3in, width=3in]{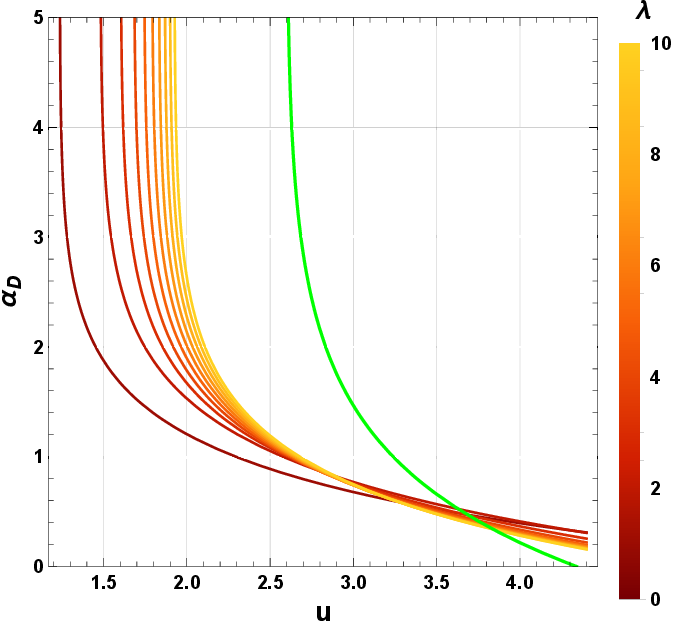}
		\end{subfigure}
		\caption{(a) Behavior of the strong deflection angle $\alpha_D$ with EH parameters $\lambda$ and charge parameter Q, keeping the other parameters fixed. (b) The strong deflection angle $\alpha_D$ is significantly influenced by the combined modifications from $F(R)$ gravity, Euler-Heisenberg electrodynamics (EH), and Gravity's Rainbow. (c) Strong deflection angle $\alpha_D$ as a function of impact parameter u, where  $\alpha_D \rightarrow \infty$ as $u \rightarrow u_{ph}$}.
		\label{fig6}
\end{figure*}

\begin{table*}
\begin{center}
 \caption{The strong lensing observables for Supermassive BHs $M87^{*}$, $SgrA^{*}$, and $NGC4649$ are estimated for different charged parameter values ( $Q=0.1, 0.2, 0.3, 0.4$) and EH parameter's values ($\lambda=0,1,6,12$) within a $F(R)$-EH Rainbow BH framework.The observable quantity $r_{mag}$ is not influenced by the BH's mass or its distance from the observer.
}\label{t2}
 \begin{tabular}{p{1cm} p{1cm} p{1.5cm} | p{1.5cm} p{1.5cm} | p{1.5cm} p{1.5cm} | p{1.5cm} p{1.5cm}  |p{1.5cm} }
 \hline
 \hline
 {Parameters} &  & & & {$ M87^*$}&   & {$ Sgr A^*$ }& NGC 4649&
\\
$\lambda$& $~$ & $Q$ & $ \theta_{\infty} (\mu as)$&$S(\mu as)$&
$\theta_{\infty} (\mu as)$ &  $S (\mu as)$ & $\theta_{\infty} (\mu as)$ &  $S (\mu as)$ &   $r_{mag}$ \\
\hline\hline
  0  &  & 0 & 19.9632 & 0.0249839 & 26.3826 & 0.0330177 & 14.6901 & 0.0183847 & 6.82188\\
\hline
   &  & 0.1 & 22.5463 & 0.0370722 & 29.7962 & 0.048993 & 16.5909 & 0.0272799  & 6.80142\\
1  &  & 0.2 & 22.1496 & 0.0385627 & 29.272  & 0.0509628 & 16.299 & 0.0283767  & 6.7368\\
   &  & 0.3 & 21.4689 & 0.041581  & 28.3724 & 0.0549517 & 15.7981 & 0.0305977 & 6.61678\\
   &  & 0.4 & 20.4648 & 0.0474194 &27.047419 &0.0626675 &15.0592  & 0.034894  & 6.41559\\
   &  & 0.5 & 19.0504 & 0.0599474 & 25.1763  & 0.079224 & 14.0184 & 0.0441128 & 6.07279\\
\hline

   &  & 0.1 & 22.5463 & 0.0370696 & 29.7963 & 0.0489826 & 16.5909 & 0.027278 & 6.80152 \\
6  &  & 0.2 & 22.1508 & 0.0385141 & 29.2736 & 0.0508986 & 16.2999 & 0.0283409& 6.73869 \\
   &  & 0.3 & 21.4759 & 0.0412527 & 28.3817 & 0.0545179 & 15.8033 & 0.0303562 & 6.62882\\
   &  & 0.4 & 20.4939 & 0.0457849 & 27.0839 & 0.0605075 & 15.0806 & 0.0336913 & 6.46995\\
   &  & 0.5 & 19.1596 & 0.0520454 & 25.3206 & 0.0687806 & 14.0988 & 0.0382978 & 6.29758\\
\hline

   &  & 0.1 & 22.5464 & 0.0370665 & 29.7964 & 0.0489855 & 16.591 & 0.0272757  & 6.80164\\
12 &  & 0.2 & 22.1522 & 0.038459 & 29.2754  & 0.0508217 & 16.3009 & 0.0282981 & 6.74095\\
   &  & 0.3 & 21.4843 & 0.0408655 & 28.3928 & 0.0540062 & 15.8094 & 0.0300713 & 6.6431\\
   &  & 0.4 & 20.5281 & 0.0439665 & 27.1291 & 0.0581043 & 15.1058 & 0.0323531 & 6.53195\\
   &  & 0.5 & 19.2789 & 0.0449542 & 25.4782 & 0.0594096 & 14.1865 & 0.03308   & 6.51865\\
\hline
\hline
\end{tabular}
\end{center}
\end{table*}
Strong lensing coefficients $\bar{a}$ and $\bar{b}$ are estimated using various values of the charge parameter $Q=0,0.1,0.2,0.3,0.4,0.5$ and the EH parameter $\lambda=0,1,6,12$ for the $F(R)$-EH gravity's Rainbow, which are assessed numerically and shown in Table \ref{t1}. At $Q=\lambda=0=R_0,f_{\epsilon}=g_{\epsilon}=1$, we discover that the Schwarzchild BH scenario is represented by $\bar{a}=1$ and $\bar{b}=-0.40023$ \cite{Bozza:2002zj}, the strong lensing coefficients.

Strong lensing coefficients $\bar{a}$ and $\bar{b}$ are shown in Fig.~\ref{fig4} and Fig.~\ref{fig5}, respectively, and are numerically estimated in Table~\ref{t1}. Figs.~\ref{fig4a} and \ref{fig5a} illustrate the variation of the strong lensing coefficients $\bar{a}$ and $\bar{b}$ as a function of two parameters: the EH parameter $\lambda$ and the charge parameter $Q$. In contrast, Figs.~\ref{fig4b} and \ref{fig5b} demonstrate the dependence of the strong lensing coefficients $\bar{a}$ and $\bar{b}$ on three parameters: the EH parameter $\lambda$, the $F(R)$ gravity parameter $f_{R_0}$, and the Rainbow parameter $f_{\epsilon}$, highlighting how the lensing coefficients are influenced by the combined modifications from $F(R)$ gravity, EH electrodynamics, and Rainbow Gravity.
For a fixed value of the EH parameter $\lambda$, while keeping all other parameters constant, Fig.~\ref{fig4a} clearly shows that the lensing coefficient $\bar{a}$ increases with increasing charge parameter $Q$. Conversely, for a fixed value of $Q$, the lensing coefficient $\bar{a}$ decreases as $\lambda$ increases. However, the lensing coefficient $\bar{b}$ increases for certain values of $\lambda$ and decreases for others, as shown in Fig.~\ref{fig5a}.
The behavior depicted in Figs.~\ref{fig4b} and \ref{fig5b} motivates the simultaneous consideration of $F(R)$ gravity, Euler–Heisenberg electrodynamics, and Rainbow Gravity, as their combined effects yield nontrivial modifications to the values of the lensing coefficients $\bar{a}$ and $\bar{b}$.

Behaviour of the strong deflection angle  $\alpha_{D}$ is shown in Fig.~\ref{fig6}. Fig.~\ref{fig6a}  illustrates the variation of the strong deflection angle  $\alpha_{D}$ as a function of two parameters: the EH parameter $\lambda$ and the charge parameter $Q$. In contrast, Fig.~\ref{fig6b}  demonstrates the dependence of the strong deflection angle  $\alpha_{D}$ on three parameters: the EH parameter $\lambda$, the $F(R)$ gravity parameter $f_{R_0}$, and the Rainbow parameter $f_{\epsilon}$, highlighting how the rong deflection angle  $\alpha_{D}$ is influenced by the combined modifications from $F(R)$ gravity, EH electrodynamics, and Rainbow Gravity.
For a fixed value of the EH parameter $\lambda$, while keeping all other parameters constant, Fig.~\ref{fig6a} clearly shows that the lerong deflection angle  $\alpha_{D}$ increases with increasing charge parameter $Q$. Conversely, for a fixed value of $Q$, the rong deflection angle  $\alpha_{D}$ decreases as $\lambda$ increases. 
The behavior depicted in Fig.~\ref{fig6b} motivates the simultaneous consideration of $F(R)$ gravity, Euler–Heisenberg electrodynamics, and Rainbow Gravity, as their combined effects yield nontrivial modifications to the srong deflection angle  $\alpha_{D}$ measurements. Fig.~\ref{fig6c}, illustrates that the strong deflection angle $\alpha_D$ as a function of impact parameter u, where  $\alpha_D \rightarrow \infty$ as $u \rightarrow u_{ph}$.

\subsection{Strong lensing observables}
Regarding the $F(R)$-EH Rainbow BH spacetime, we then investigate the observable characteristics of strong gravitational lensing. We suppose that the source is located in a flat spacetime and that the observer is almost parallel to the optical axis. The lens equation can be expressed as follows in asymptotically flat spacetime when the source is directly behind a BH:\cite{bozza2001strong,islam2020gravitational,bozza2008comparison} 
\begin{equation}\label{28}
\psi=\theta-\frac{D_{LS}}{D_{OS}}\Delta\alpha_n.
\end{equation}
The deflection offset,$\Delta\alpha_n$, is given by
\begin{equation}
    \Delta\alpha_n= \alpha_{D}(\theta) - 2n\pi,
\end{equation}
where the number of loops is denoted by the integer n. Here, $\theta$ and $\psi$ stand for the angular separations between the observer-source and the source-BH (lens), respectively.
The distances between the observer lens, lens-source, and observer-source are denoted as $D_{OL}$, $D_{LS}$, and $D_{OS}$, respectively:

\begin{equation}
    D_{OS}=D_{OL}+D_{LS}
\end{equation} 
Equations (\ref{24}) and (\ref{28}) provide a representation of the angle separation between the BH (lens) and the $n^{th}$ relativistic image as:

\begin{equation}\label{31}
\theta_n=\theta_n^0+\frac{u_{ph}e_n(\psi-\theta_n^0)D_{OS}}{\bar{a}D_{LS}D_{OL}}
\end{equation}
Where
\begin{equation}
e_n=\exp\left(\frac{\bar{b}-2n\pi}{\bar{a}}\right)
\end{equation}
and
\begin{equation}
    \theta_n^0=\frac{u_{ph}(1-e_{n})}{D_{OL}}
\end{equation}
In this case,$\theta_n^0$ indicates the image's angular position after a complete of $2n\pi$ full rotations of the photon around the BH (lens). Equation (\ref{31}) makes it evident that when $\psi=\theta_n^0$, the position of the image and the source coincide.
 This indicates that the source and the image are on the same side when $\theta_n=\theta_n^0$. 
The magnification of the relativistic effect in strong gravitational lensing is determined by the ratio of the solid angle covered by the $n$th image to the source \cite{drvirbhaa2000schwarzschild}.
The $n$th relativistic image's magnification can be found as follows \cite{Virbhadra:1999nm,Bozza:2002zj,PhysRevD.62.084003}:
\begin{equation}
\mu_n=\left.\left(\frac{\psi}{\theta}\frac{d\psi}{d\theta}\right)^{-1}\right|_{\theta_n^0}=\frac{{u_{ph}}^2e_n(1+e_n)D_{OS}}{\bar{a}\psi D_{LS}{D_{OL}}^2},
\label{flux}
\end{equation}
As n rises, the magnification of the initial relativistic image decreases exponentially, making it the brightest.
Relativistic images are typically faint due to the inverse relationship between magnification and the square of the distance between the observer and the lens.
The Einstein ring, which forms in the limit as $\psi \to 0$, represents perfect alignment. The study investigates the significant gravitational lensing effects by computing three key lensing observables. The brightest image, $\theta_{1}$, is recognized as a separate entity, while other inner images are clustered near $\theta_{\infty}$(where $\theta_{n}|_{n \to \infty}=\theta_{\infty}$)\cite{Bozza:2002zj}. The deflection angle (\ref{24}) and lens Eq. (\ref{28}) can be used to calculate strong lensing observables, which include the angular position of the set of images $\theta_{\infty}$, the angular separation $S$ between outer and inner images, and the relative magnification $r_{mag}$ between them.
 These three quantities can be accurately described using the formulas shown in: \cite{Bozza:2002zj, Kumar:2022fqo}
\begin{equation}\label{35}
    \theta_{\infty}=\frac{u_{ph}}{D_{OL}}
\end{equation}
\begin{equation}\label{36}
S= \theta_1-\theta_{\infty}\approx\theta_{\infty}e^\frac{(\bar{b} -2\pi)}{\bar{a}}
\end{equation}
\begin{equation}\label{37}
    \begin{split}
    r_{mag}=\frac{\mu_1}{\sum_{n=2}^{\infty}\mu_n}\approx&2.5\log_{10}\left[\exp\left(\frac{2\pi}{\bar{a}}\right)\right]\\
    &=\frac{5\pi}{\bar{a}log{10}}
    \end{split}
\end{equation}
The lensing coefficients $\bar{a}$, $\bar{b}$, and the minimum impact parameter $u_{ph}$ can be determined by solving Eqs. (\ref{35}),(\ref{36}) and  (\ref{37}) in reverse if we quantify the strong lensing observables $\theta_{\infty}$,$S$ and $r_{mag}$ through observations. The observed data can be compared with the theoretically obtained values. The findings simplify the distinction between RN BH, Schwarzchild BH, and $F(R)$-Euler-Heinsenberg Rainbow BH.

\begin{figure*}[htbp]
 \captionsetup[subfigure]{labelformat=simple}
    \renewcommand{\thesubfigure}{(\alph{subfigure})}
		\begin{subfigure}{.45\textwidth}
			\caption{}\label{fig7a}
			\includegraphics[height=3in, width=3in]{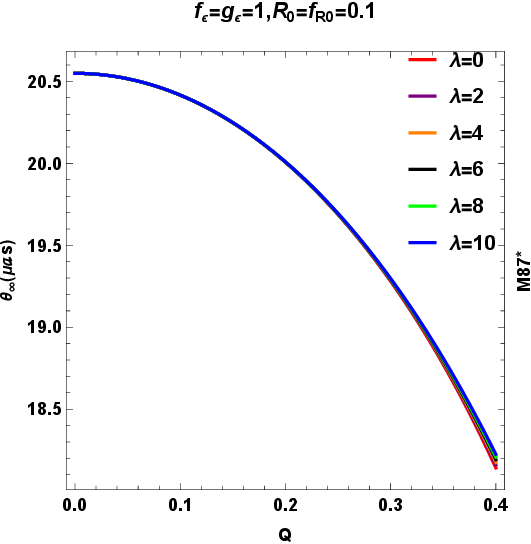}
		\end{subfigure}
            \begin{subfigure}{.4\textwidth}
			\caption{}\label{fig7b}
			\includegraphics[height=3in, width=3in]{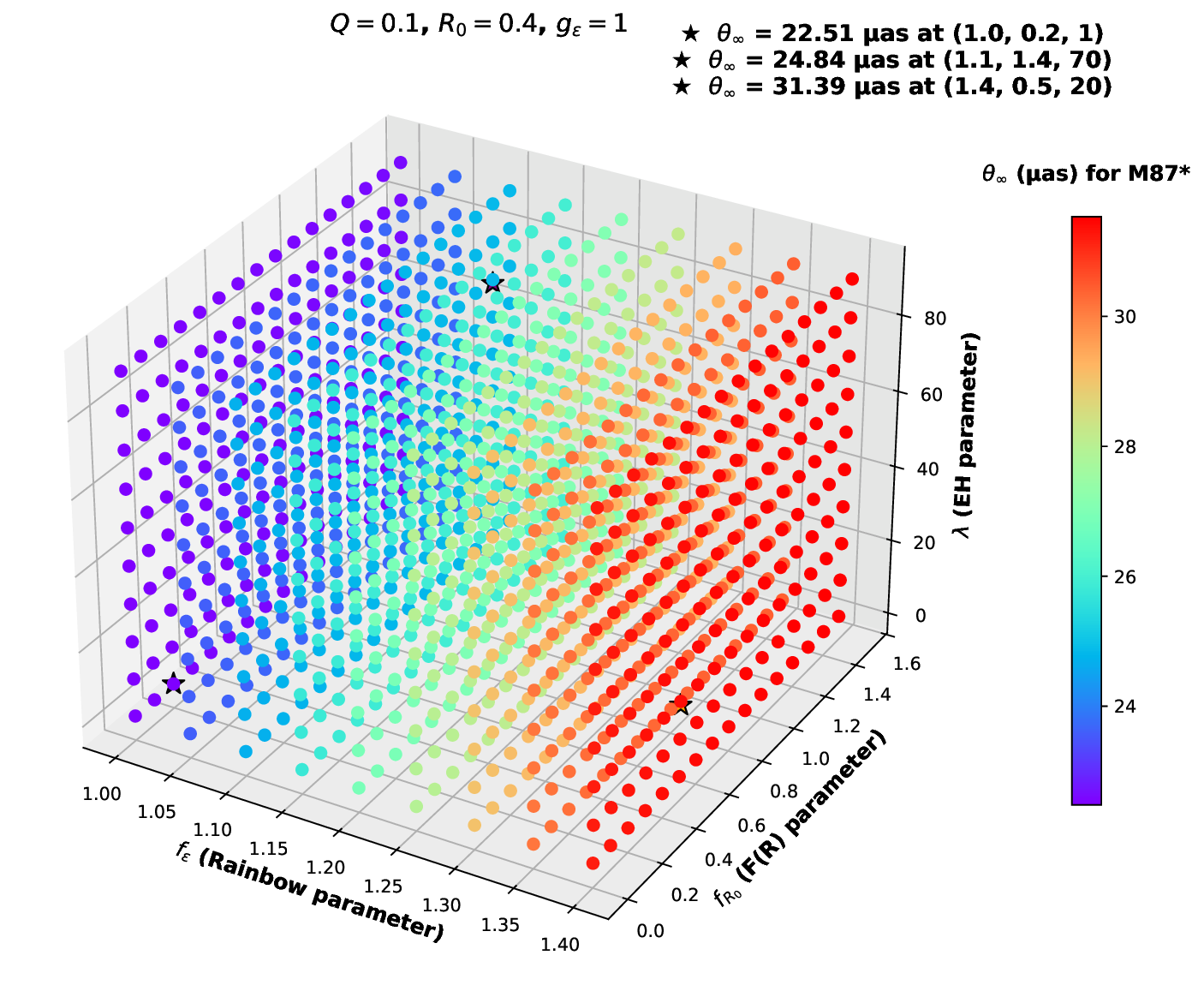}
		\end{subfigure}
  
		\caption{(a) Behavior of the angular position $\theta_{\infty} (\mu as)$ for M87$^*$ with EH parameters $\lambda$ and charge parameter Q, keeping the other parameters fixed. (b) The angular position $\theta_{\infty} (\mu as)$ for M87$^*$ is significantly influenced by the combined modifications from $F(R)$ gravity, Euler-Heisenberg electrodynamics (EH), and Gravity's Rainbow.}
		\label{fig7}
\end{figure*} 

\begin{figure*}[htbp]
 \captionsetup[subfigure]{labelformat=simple}
    \renewcommand{\thesubfigure}{(\alph{subfigure})}
		\begin{subfigure}{.45\textwidth}
			\caption{}\label{fig8a}
			\includegraphics[height=3in, width=3in]{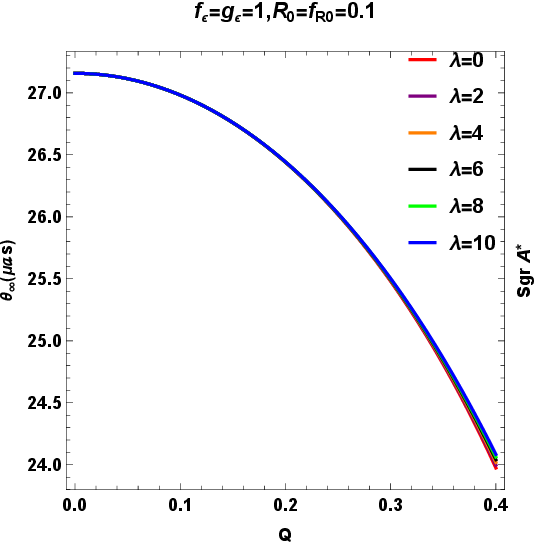}
		\end{subfigure}
            \begin{subfigure}{.4\textwidth}
			\caption{}\label{fig8b}
			\includegraphics[height=3in, width=3in]{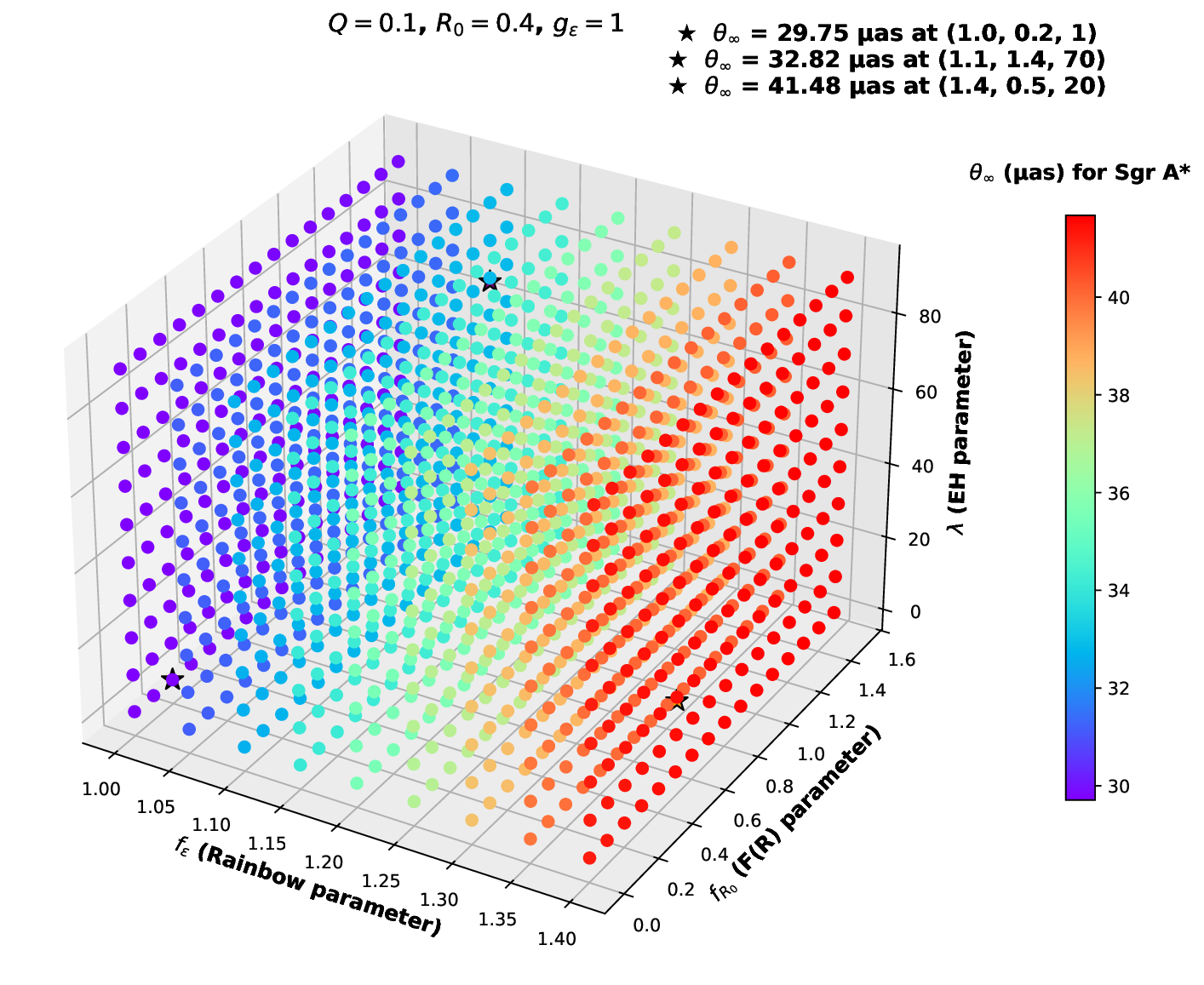}
		\end{subfigure}
  
		\caption{(a) Behavior of the angular position $\theta_{\infty} (\mu as)$ for SgrA$^*$ with EH parameters $\lambda$ and charge parameter Q, keeping the other parameters fixed. (b) The angular position $\theta_{\infty} (\mu as)$ for SgrA$^*$ is significantly influenced by the combined modifications from $F(R)$ gravity, Euler-Heisenberg electrodynamics (EH), and Gravity's Rainbow. }
		\label{fig8}
\end{figure*}

\begin{figure*}[htbp]
 \captionsetup[subfigure]{labelformat=simple}
    \renewcommand{\thesubfigure}{(\alph{subfigure})}
		\begin{subfigure}{.45\textwidth}
			\caption{}\label{fig9a}
			\includegraphics[height=3in, width=3in]{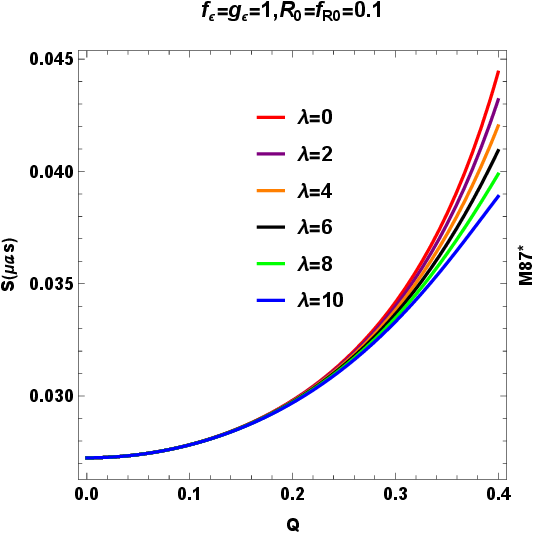}
		\end{subfigure}
            \begin{subfigure}{.4\textwidth}
			\caption{}\label{fig9b}
			\includegraphics[height=3in, width=3in]{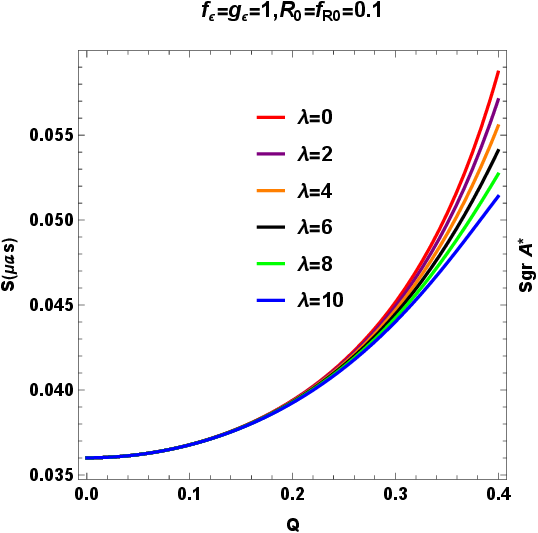}
		\end{subfigure}
  
		\caption{ Behavior of the angular separation $S (\mu as)$ for M87$^*$ with EH parameters $\lambda$ and charge parameter Q, keeping the other parameters fixed. (b)  Behavior of the angular separation $S (\mu as)$ for Sgr A$^*$ with EH parameters $\lambda$ and charge parameter Q, keeping the other parameters fixed.  }
		\label{fig9}
\end{figure*}

\subsubsection{Lensing observables and the EH Rainbow BH parameters:}
With the supermassive BHs $M87^{*}$, $SgrA^{*}$, and $NGC 4649$ in the cores of nearby galaxies taken into consideration,  observables variables $\theta_{\infty}$, $S$, and $r_{mag}$ are determined by numerical simulations, for the EH Rainbow gravity (see Table \ref{t2}). The mass of $M87^{*}$ is about $6.5\times10^9M_{\odot}$, and it is $16.8$ Mpc away from Earth according to \cite{event2019first}.The mass of $SgrA^{*}$, according to data from \cite{gillessen2017update}, is roughly $4.28\times10^6M_{\odot}$, and the distance is roughly $8.32$ kpc. According to \cite{kormendy2013coevolution}, the mass of $NGC4649$ is around $4.72\times10^9M_{\odot}$, and the distance is roughly $16.46$ Mpc.

\begin{figure*}[htbp]
 \captionsetup[subfigure]{labelformat=simple}
    \renewcommand{\thesubfigure}{(\alph{subfigure})}
		\begin{subfigure}{.45\textwidth}
			\caption{}\label{fig10a}
			\includegraphics[height=3in, width=3in]{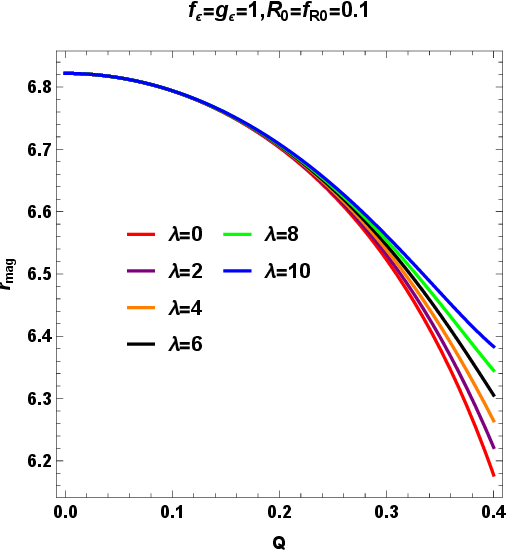}
		\end{subfigure}
            \begin{subfigure}{.4\textwidth}
			\caption{}\label{fig10b}
			\includegraphics[height=3in, width=3in]{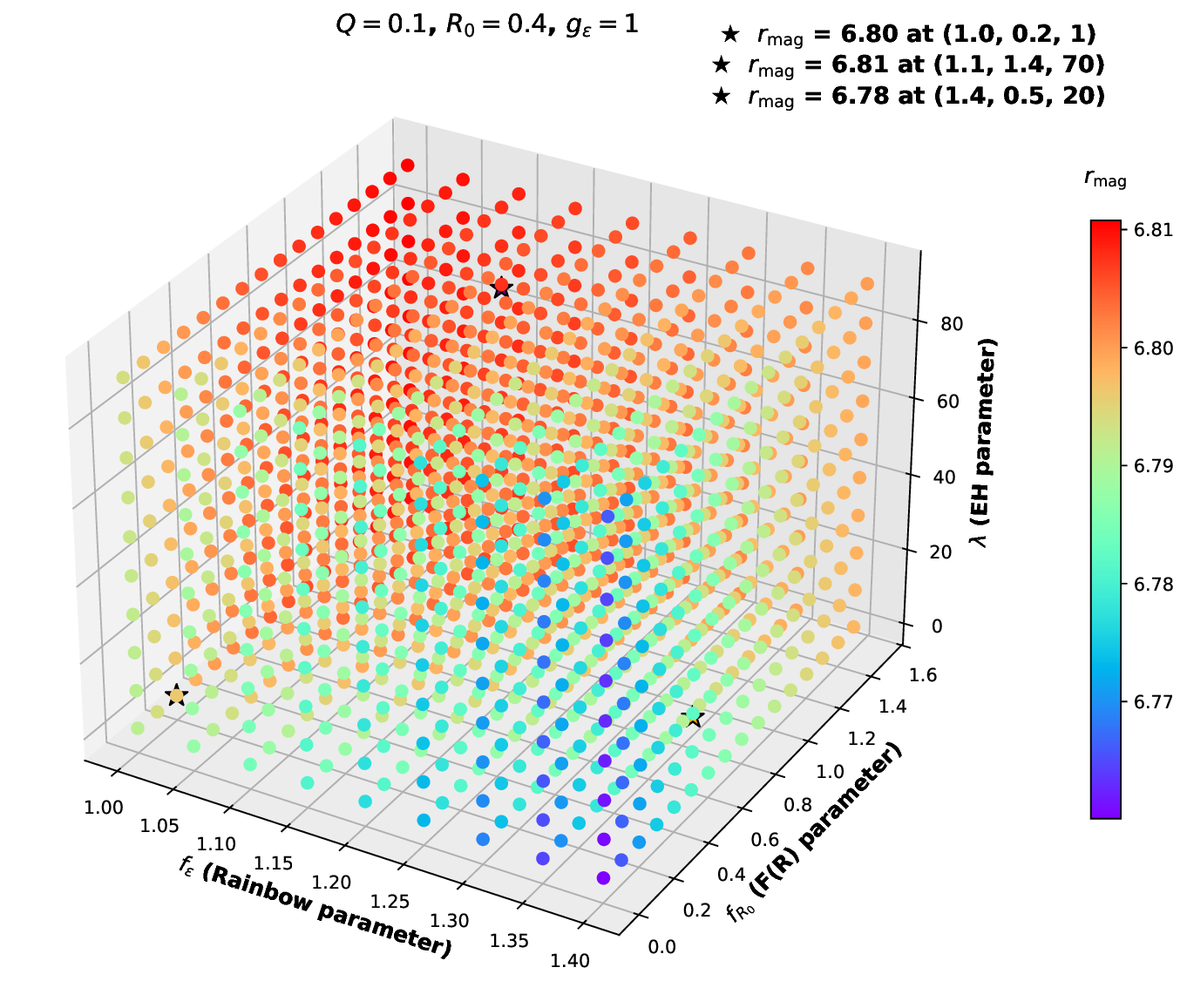}
		\end{subfigure}
  
		\caption{(a) Behavior of the relative magnification $r_{mag}$ with EH parameters $\lambda$ and charge parameter Q, keeping the other parameters fixed. (b) The relative magnification $r_{mag}$ is significantly influenced by the combined modifications from $F(R)$ gravity, Euler-Heisenberg electrodynamics (EH), and Gravity's Rainbow.  }
		\label{fig10}
\end{figure*}

 Lensing observable quantities $\theta_{\infty}$, $S$, and $r_{mag}$ for supermassive BHs $M87^{*}$ and $SgrA^{*}$ are depicted in Figs.~\ref{fig7},\ref{fig8};\ref{fig9}, and \ref{fig10} respectively  and numerically estimated in Table \ref{t2}. Figs.~\ref{fig7a}\& \ref{fig8a} illustrate the variation of the angular position $\theta_{\infty}$ as a function of two parameters: the EH parameter $\lambda$ and the charge parameter $Q$. In contrast, Figs.~\ref{fig7b}\&~\ref{fig7b} demonstrate the dependence of $\theta_{\infty}$ on three parameters: the EH parameter $\lambda$, the $F(R)$ gravity parameter $f_{R_0}$, and the Rainbow parameter $f_{\epsilon}$, highlighting how $\theta_{\infty}$ is influenced by the combined modifications from $F(R)$ gravity, EH electrodynamics, and Rainbow Gravity.
For a fixed value of the EH parameter $\lambda$, while keeping all other parameters constant, Figs.~\ref{fig7a}\& \ref{fig8a} clearly show that the  angular position $\theta_{\infty}$ decreases with increasing charge parameter $Q$. Conversely, for a fixed value of $Q$, the angular position $\theta_{\infty}$ increases as $\lambda$ increases. Notably, when $Q = 0 = R_0$ and $f_{\epsilon}=g_\epsilon=1$, the angular position $\theta_{\infty}$ reduces to $\theta_{\infty}= 19.9632$ for $M87^*$ and $\theta_{\infty}= 26.3826$ for $Sgr A^*$, which corresponds to the Schwarzschild BH, consistent with the result in \cite{Bozza:2002zj}.
The behavior depicted in Figs.~\ref{fig7b}\&~\ref{fig8b} motivates the simultaneous consideration of $F(R)$ gravity, Euler–Heisenberg electrodynamics, and Rainbow Gravity, as their combined effects yield nontrivial modifications to the angular position $\theta_{\infty}$ value.
Fig.\ref{fig9} depicts the behavior of observable quantitiy $S$ for supermassive BHs $M87^{*}$ and $SgrA^{*}$.
Figs.\ref{fig9a} \ref{fig9b} show that for a fixed value of the EH parameter $\lambda$, the observable quantity  $S$ increases with increasing charge parameter Q and for the fixed value of the charged parameter Q, with increasing value of the EH parameter $\lambda$, the observable quantity $S$ decreases for both supermassive BHs $M87^{*}$ and $SgrA^{*}$. The value of the observable quantity S, very small for both supermassive BHs $M87^{*}$ and $SgrA^{*}$. This suggests that all of the relativistic images in this situation are densely grouped for $F(R)$-EH Rainbow BH. However, as the EH parameter $\lambda$ increases, the observable quantity $r_{mag}$, which represents the relative magnification, shows an increasing trend (see Fig.\ref{fig10a}). Furthermore, it decreases when the charged parameter Q increases for a fixed value of $\lambda$. The behavior depicted in Fig.~\ref{fig10b} motivates the simultaneous consideration of $F(R)$ gravity, Euler–Heisenberg electrodynamics, and Rainbow Gravity, as their combined effects yield nontrivial modifications to the the relative magnification $r_{mag}$ value.\\

In our calculations (see Table.\ref{t2}), taking into account the same mass and
distance, as demonstrated by $M87^{*}$ for $F(R)$-EH Rainbow BH, The angular position of innermost image $\theta_{\infty}$, is found to be higher except $Q=0.5$ than the case of both  Schwarzschild and RN BHs scenario.The angular image separation between outermost $(\theta_1)$ and innermost packed images ($\theta_{\infty}$) of relativistic images, represented by the symbol $S$, tends towards zero in the context of $F(R)$-EH Rainbow BH, higher than in RN and Schwarzschild BHs. Additionally, the $F(R)$-EH Rainbow BH has a smaller relative magnification $r_{mag}$. According to these results, the outermost images of the $F(R)$-EH Rainbow BH are detectable and densely packed with the innermost images, in contrast to the other images of the BH. Moreover, the most recent technology offers the ability to distinguish between the $F(R)$-EH Rainbow BH and a regular Schwarzschild BH or RN BH by resolving the outermost relativistic image. When $\lambda=12$ and $Q= 0.1$ for the RN BH, the differences in their values are approximately 3.13 $\mu as$ and 0.1 in magnitude, respectively, and the differences of the Schwarzschild BH are approximately 2.58 $\mu as$ and 0.02 in magnitude, respectively. Current technology (EHT) struggles to distinguish the $F(R)$-EH Rainbow BH from regular BH like Schwarzschild or RN BH. As a result, making such observations is quite difficult, and we might have to wait for the next generation of EHT (ngEHT) to make improvements.

\subsection{Einstein Ring}
\begin{figure*}[htbp]
\begin{center}	
\begin{tabular}{p{9cm} p{9cm}}
\includegraphics[scale=.8]{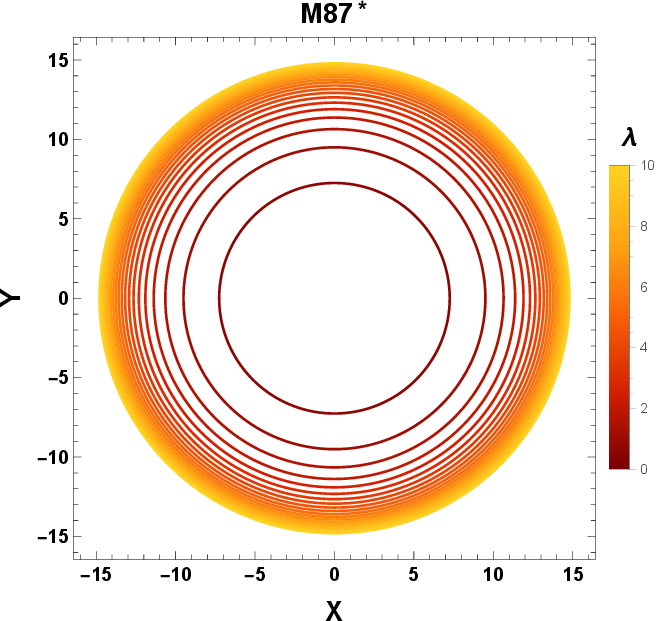}&
\includegraphics[scale=.8]{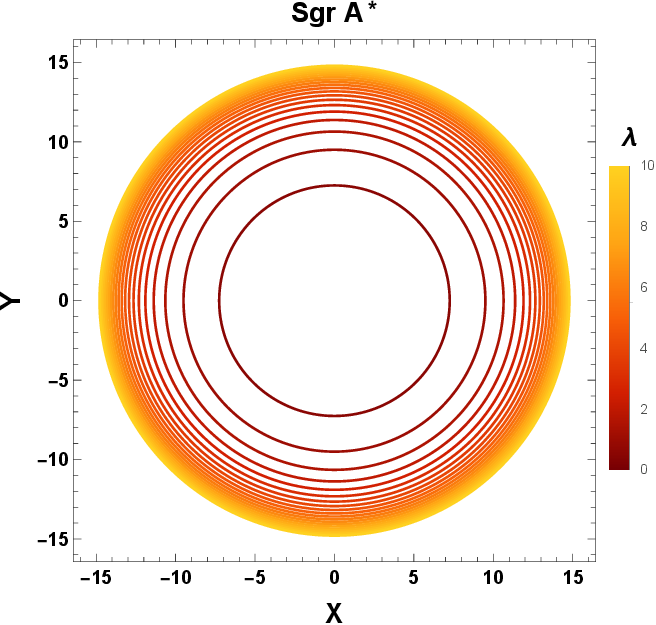}
\end{tabular}
	\caption{Behavior of the angular radius $\theta_{1}^{E}$ for the supermassive BHs $M87^{*}$ (left panel) and $SgrA^{*}$ (right panel) with parameter $\lambda$.} \label{fig11}		
\end{center}   
\end{figure*}

The BH bends light rays in all directions when the observer, BH (lens), and source are precisely aligned ($\psi=0$), creating a ring-shaped image. This phenomenon, which has been thoroughly investigated in numerous research studies, including \cite{einstein1936lens,liebes1964gravitational,mellier1999probing,bartelmann2001weak,schmidt2008weak,guzik2010tests}, is frequently called an Einstein ring.The angular radius of the $n^{th}$ relativistic image can be obtained by simplifying Eq. (\ref{31}) for $\psi=0$ as:
\begin{equation}
    \theta_{n}=\theta_{n}^{0}\Big(1-\frac{u_{ph}e_{n}D_{OS}}{\bar{a}D_{LS}D_{OL}}\Big)
\end{equation}
Under the assumption that $D_{OL}>>u_{ph}$, By placing the BH (lens) halfway between the source and the observer, the angular radius of the $n$th relativistic Einstein ring in the spacetime of a $F(R)$-EH Rainbow BH may be written as follows: $D_{OS}=2D_{OL}$
\begin{equation}
    \theta_{n}^{E}=\frac{u_{ph}(1+e_{n})}{D_{OL}}
\end{equation}
The outermost Einstein ring is represented by the angular radius $\theta_{1}^{E}$, as seen in Fig.\ref{fig11} (left and right panels) for $M87^{*}$ and $SgrA^{*}$, the supermassive BH, respectively. It can be seen from these two figures that as the parameter $\lambda$ increases, the angular radius $\theta_{1}^E$ of the outermost Einstein ring due to $F(R)$-EH Rainbow BH is also increases.

\subsection{ Time Delay in the Strong Field Limit}
The time delay, which represents the time difference between the formation of two relativistic images, is a crucial observable in the strong gravitational lensing phenomenon. The varying paths taken by photons around the BH lead to differences in their travel times, resulting in time delays between different relativistic images. This occurs because the time required for photons to traverse these distinct trajectories differs. The time delay between two relativistic images can be determined by analyzing the observed time signals of those images \cite{bozza2004time}. Bozza and Manchini \cite{bozza2004time} provide an expression for the time taken by a photon to complete an orbit around the BH.
\begin{equation}\label{32}
    \tilde{T}=\tilde{a}log\Big(\frac{u}{u_{ph}}-1\Big)+\tilde{b}+\mathcal{O}(u-u_{ph}).
\end{equation}
One can determine the time difference between two relativistic images using the previously given Eq. (\ref{32}). The formula determines the time delay between two relativistic images on the same side of a BH in a spherically symmetric BH spacetime:

\begin{equation}\label{33}
    \Delta T_{2,1}=2\pi u_{ph}=2\pi D_{OL}\theta_{\infty}
\end{equation}
We have calculated the time delays for the $F(R)$-EH Rainbow BH, RN BH, and Schwarzschild BH using Eq.~(\ref{33}), by considering several supermassive BHs. Table~\ref{t3} presents the results for several supermassive BHs located at the centers of neighboring galaxies. Table~\ref{t3} shows that the time delays for the $F(R)$-EH Rainbow BH are greater than those for the Schwarzschild BH and the RN BH. For example, for NGC 4751, the time delay for the $F(R)$-EH Rainbow BH ($\sim 7018.07$ min) is significantly greater than that of both the Schwarzschild BH ($\sim 6523.75$ min) and the RN BH ($\sim 6100.07$ min). These results suggest that if one can resolve the two different relativistic images, the time difference between them might provide a good opportunity to distinguish the $F(R)$-EH Rainbow BH from other ordinary BHs, such as the RN BH and the Schwarzschild BH.
 \begin{table*}
 \caption{Time delay estimate for the Schwarzschild BH, RN BH, and $F(R)$-EH Rainbow BH, respectively, for various supermassive BHs. Solar Masses and Mpc units \cite{kormendy2013coevolution} are used to express the masses and distances. Minutes are used to estimate the time delays $\Delta T_{2,1}$.
\label{t3}}
\begin{tabular}{ p{3cm}p{3cm}p{3cm}p{3cm}p{3cm}p{3cm}}
\hline
\hline
 Galaxy  &$M(M_{\odot})$& $D_{ol}(Mpc)$&
 Schwarzschild\vfill BH ($R_{0}=0=Q,f_{\epsilon}=g_{\epsilon}=1$)  &   RN BH\vfill & EH Rainbow BH\\

\hline
\hline
\multirow{22}{2em}
MMilkyway &  0.43 $\times$ $10^7$ & 0.0083&11.4968&10.7501&12.3679\\
M87  & $0.65\times 10^{10} $& 16.68&17378.8&16250.2&18695.2\\
NGC 7457 & $0.895\times 10^7 $& 12.53&23.9293&22.3753&25.7425\\
NGC 4395 & $0.36\times 10^6 $& 4.3&0.96252&0.90001&1.03545\\
NGC 2778 & $0.145\times 10^8 $&23.44&38.7682&36.2504&41.7058\\
NGC 4026 & $0.18\times 10^9 $& 13.35&481.26&450.002&517.727\\
NGC 3379 & $0.416\times 10^9 $& 10.70&1112.25&1040.01&1196.52\\
NGC 3607 & $0.137\times 10^9 $& 22.65&366.292&342.504&394.048\\
NGC 4261 & $0.529\times 10^9 $& 32.36&1414.37&1322.52&1521.54\\
NGC 4486A& $0.144\times 10^8 $& 18.36&38.5008&36.0004&41.4181\\
NGC 3842 & $0.909\times 10^{10} $& 92.2&24303.6&22725.3&26145.2\\
NGC 4649 & $0.472\times 10^{10} $& 16.46&12619.7&11800.1&13575.9\\
NGC 4751 & $0.244\times 10^{10} $& 32.81&6523.75&6100.07&7018.07\\
NGC 5516 & $0.369\times 10^{10} $& 55.3&9865.83&9225.11&10613.4\\
NGC 5576 & $0.273\times 10^9 $& 25.68&729.911&682.508&785.219\\
NGC 6251 & $0.614\times 10^9 $& 108.4&1641.63&1535.02&1766.02\\
NGC 5077 & $0.855\times 10^9 $& 38.7&2285.99&2137.52&2459.2\\
NGC 6861 & $0.210\times 10^{10} $& 28.71&5614.7&5250.06&6040.15\\
NGC 7052 & $0.396\times 10^9 $& 70.4&1058.77&990.011&1139\\
Cygnus A & $0.266\times 10^{10} $& 242.7&7111.95&6650.08&7650.85\\
NGC 7768 & $0.134\times 10^{10} $& 116.0&3582.72&3350.04&3854.19\\
\hline
\hline
\end{tabular}
\end{table*}
\section{Astrophysical Constraints the EULER-HEISENBERG parameter of the BH by EHT observations data of $M87^*$ and $SgrA^*$}\label{sec5}

\begin{figure*}[htbp]
 \captionsetup[subfigure]{labelformat=simple}
    \renewcommand{\thesubfigure}{(\alph{subfigure})}
		\begin{subfigure}{.45\textwidth}
			\caption{}\label{fig12a}
			\includegraphics[height=3in, width=3in]{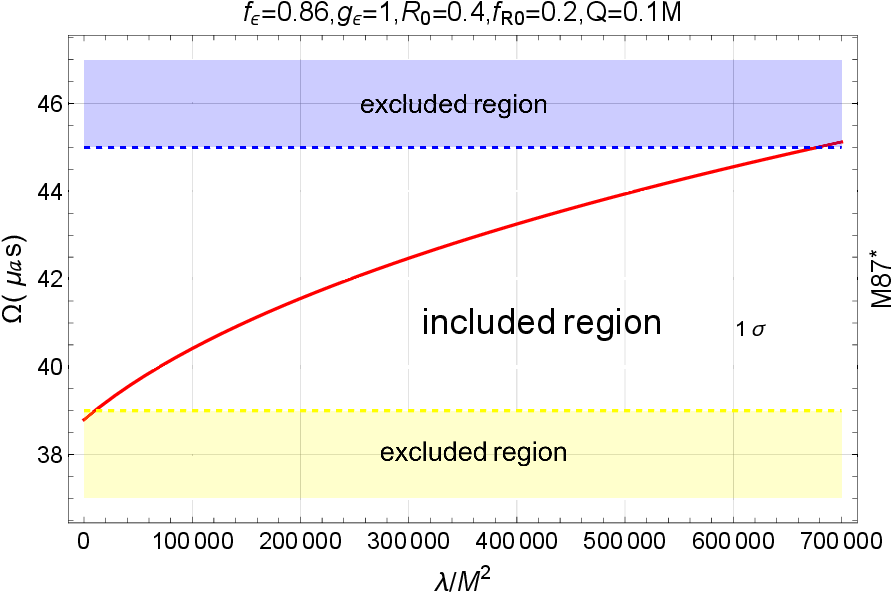}
		\end{subfigure}
            \begin{subfigure}{.4\textwidth}
			\caption{}\label{fig12b}
			\includegraphics[height=3in, width=3in]{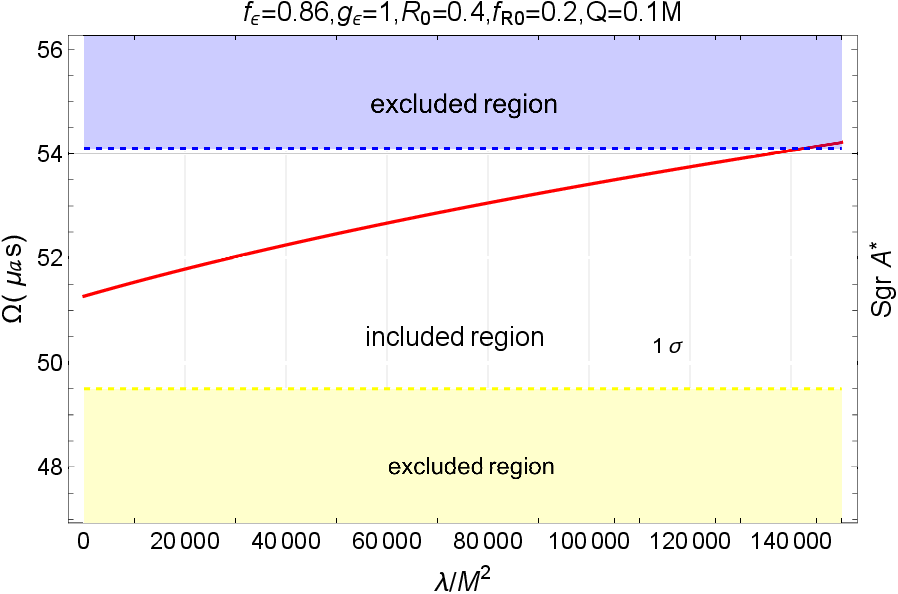}
		\end{subfigure}
  
		\caption{The angular diameter of the shadow, $\Omega = 2\theta_{\infty} (\mu as)$, is shown as a function of $ \frac{\lambda}{M^2}$ for $M87^{*}$ (left panel) and $Sgr A^{*}$ (right panel) keeping the other parameters fixed . The allowed (white) region corresponds to the parameter space that is $1\sigma$ consistent with EHT observations, while the excluded (blue and yellow) regions represent areas that are $1\sigma$ inconsistent, imposing constraints on the EH BH parameter $\lambda$.}
		\label{fig12}
\end{figure*}

Some authors have constrained individual parameters in the present BH model such as the Rainbow functions \( f_{\epsilon} \) and \( g_{\epsilon} \) in \cite{Ali:2014aba}, and the charge parameter \( Q \) in \cite{Bozza:2002zj}, among others. However, a comprehensive constraint analysis involving all these parameters simultaneously is beyond the scope of the present work. Here aims to  constraints on BH parameter  $\lambda$ only within the framework of $F(R)$-Euler-Heisenberg (EH) Rainbow gravity using recent observational results from the Event Horizon Telescope (EHT) by setting a fixed value of the other BH parameters. The EHT collaboration has provided groundbreaking images of the supermassive BHs M87$^*$ and Sgr A$^*$, offering new avenues to test alternative theories of gravity.
The EHTs first major achievement was the successful imaging of the shadow of the BH at the center of the galaxy M87$^*$ ~\cite{EventHorizonTelescope:2019dse,EventHorizonTelescope:2019uob,EventHorizonTelescope:2019jan,EventHorizonTelescope:2019ths,EventHorizonTelescope:2019pgp,EventHorizonTelescope:2019ggy}. These observations revealed strong gravitational lensing effects and enabled the study of the near-horizon geometry. M87$^*$ is located approximately 16.8 Mpc away, with a measured mass of $(6.5 \pm 0.7) \times 10^9 M_{\odot}$. The observed angular diameter of its shadow is $\Omega = 42 \pm 3  \mu as$.
In 2022, the EHT collaboration released the first image of Sgr A$^*$, the supermassive BH at the center of the Milky Way galaxy. This BH has a mass of $M = (4.3 \pm 0.7) \times 10^6 M_{\odot}$, a distance of $d = 8.35$ kpc, and an angular shadow diameter measured to be 
$51.8 \pm 2.3 $ $\mu as$~\cite{EventHorizonTelescope:2022wkp}.
Using the observable $\theta_{\infty}$, which represents the angular radius of the BH shadow due to strong lensing, we impose constraints on the parameter $\lambda/M^2$ characterizing the $F(R)$ EH Rainbow gravity model: For M87$^*$ (see Fig.~\ref{fig12a}), the allowed range is $0 \leq \lambda/M^2 \leq 6.86 \times 10^5$. For Sgr A$^*$ (see Fig.~\ref{fig12b}), the constraint is $0 \leq \lambda/M^2 \leq 1.45 \times 10^5$.
These ranges of the parameters are obtained based on the choice of other parameters. They change with different choices of parameter sets. These findings demonstrate that the $F(R)$-EH Rainbow gravity framework can accommodate the observed shadow properties of both M87$^*$ and Sgr A$^*$, providing a viable alternative to General Relativity in the strong-field regime. Future high-precision observations may further tighten these bounds and offer deeper insights into the underlying gravitational theory.

\section{Observational Signatures of Gravitational Lensing in Different Rainbow Function Frameworks}\label{sec6}
The robustness of the gravitational lensing predictions under different choices of Rainbow functions is indeed crucial to validate the generality of our results. Here, we  investigate by conducting a comparative analysis using three well-known and phenomenologically motivated Rainbow function models \cite{Ali:2014aba}.

\begin{itemize}
   \item \textbf{Model~I~ (MDR1):} Proposed by Amelino-Camelia et al.~\cite{Amelino-Camelia:1996bln,Amelino-Camelia:2008aez}, this model is inspired by loop quantum gravity and noncommutative spacetime. The Rainbow functions are:
    \begin{equation}
    f_{\epsilon}(E/E_p) = 1, \quad g_{\epsilon}(E/E_p) = \sqrt{1 - \eta \frac{E}{E_p}},
    \label{eq:mdr1}
    \end{equation}
    where \( \eta \) is a dimensionless parameter. This form preserves time reparameterization symmetry while modifying spatial geometry.
~~~~~~~~~~~~~~~~~~~~~~~~~~~
    \item \textbf{Model ~II ~(MDR2):}  Also introduced by Amelino-Camelia et al.~\cite{Amelino-Camelia:1997ieq}, this form was motivated by high-energy gamma-ray observations from cosmological sources:
    \begin{equation}
    f_{\epsilon}(E/E_p) = \frac{e^{\alpha E/E_p} - 1}{\alpha E/E_p}, \quad g_{\epsilon}(E/E_p) = 1,
    \label{eq:mdr2}
    \end{equation}
    where \( \alpha \) is a constant of order unity. Here, the time component of the metric is modified, affecting the gravitational redshift and time delay.
    \item \textbf{Model~III~ (MDR3):}  Proposed by Magueijo and Smolin~\cite{Magueijo:2001cr}, this model assumes a modified dispersion relation that maintains the constancy of the speed of light:
    \begin{equation}
    f_{\epsilon}(E/E_p) = g_{\epsilon}(E/E_p) = \frac{1}{1 - \gamma  E/E_p},
    \label{eq:mdr3}
    \end{equation}
    where $\gamma$ is a dimensionless parameter. Both temporal and spatial components are modified, and this model is often used in studies of BH thermodynamics and cosmology.
\end{itemize}

The parameters \( \eta \), \( \alpha \), and \( \gamma \) are typically assumed to be order of unity, which implies significant Rainbow corrections only near the Planck scale. However, if these parameters are smaller, they may signal the existence of an intermediate fundamental scale. To remain consistent with current observational constraints, the value of these parameters becomes $\eta, \alpha, \gamma \leq 10^{17}$ \cite{Ali:2014aba}.

\begin{figure*}[htbp]
 \captionsetup[subfigure]{labelformat=simple}
    \renewcommand{\thesubfigure}{(\alph{subfigure})}
		\begin{subfigure}{.45\textwidth}
			\caption{}\label{13a}
			\includegraphics[height=3in, width=3in]{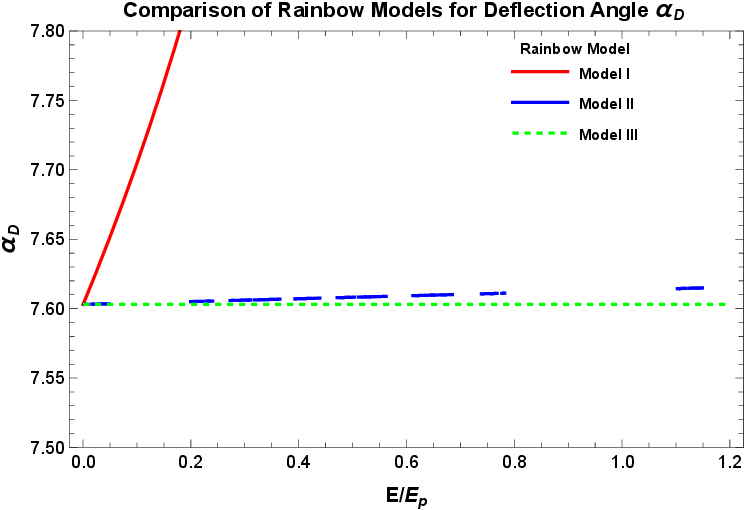}
		\end{subfigure}
		\caption{(a) Variation of the strong deflection angle $\alpha_{D}$ for different functional forms of the Rainbow functions, with all other parameters held fixed.} 
		\label{fig13}
\end{figure*}

\begin{figure*}[htbp]
 \captionsetup[subfigure]{labelformat=simple}
    \renewcommand{\thesubfigure}{(\alph{subfigure})}
		\begin{subfigure}{.45\textwidth}
			\caption{}\label{14a}
			\includegraphics[height=3in, width=3in]{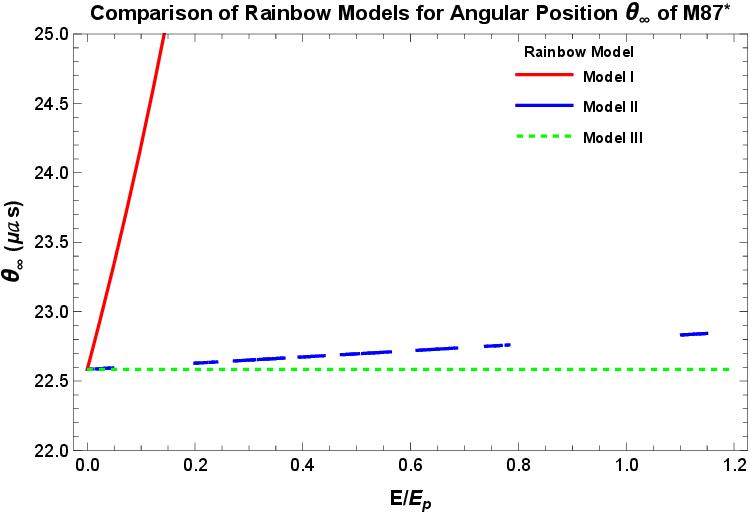}
		\end{subfigure}
            \begin{subfigure}{.4\textwidth}
			\caption{}\label{14b}
			\includegraphics[height=3in, width=3in]{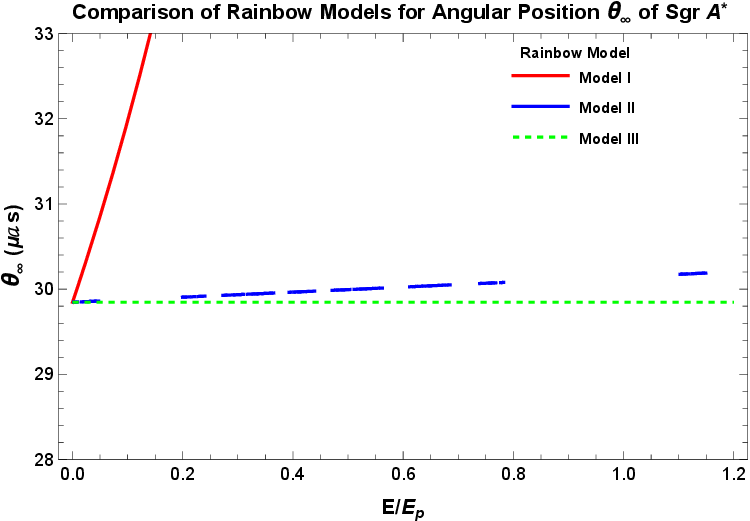}
		\end{subfigure}
  
		\caption{(a) Variation of the angular position $\theta_{\infty} (\mu as)$ for different functional forms of the Rainbow functions, with all other parameters held fixed, in the context of M87$^*$. (b) Variation of the angular position $\theta_{\infty} (\mu as)$ for different functional forms of the Rainbow functions, with all other parameters held fixed, in the context of Sgr A$^*$.}
		\label{14}
\end{figure*}

\begin{figure*}[htbp]
 \captionsetup[subfigure]{labelformat=simple}
    \renewcommand{\thesubfigure}{(\alph{subfigure})}
		\begin{subfigure}{.45\textwidth}
			\caption{}\label{15a}
			\includegraphics[height=3in, width=3in]{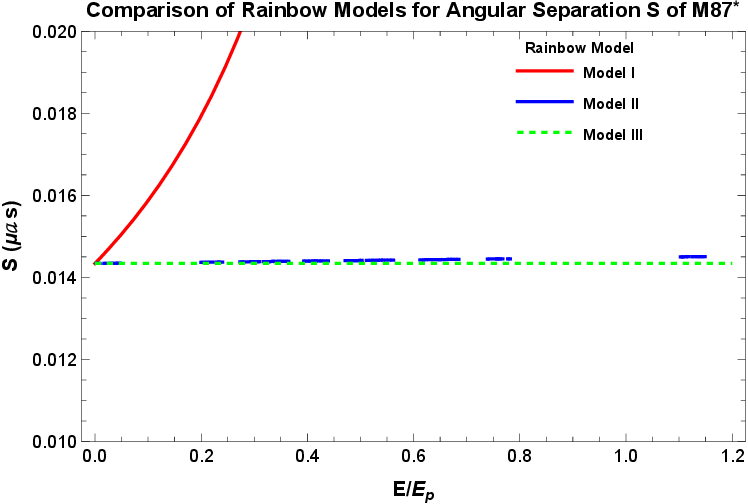}
		\end{subfigure}
            \begin{subfigure}{.4\textwidth}
			\caption{}\label{15b}
			\includegraphics[height=3in, width=3in]{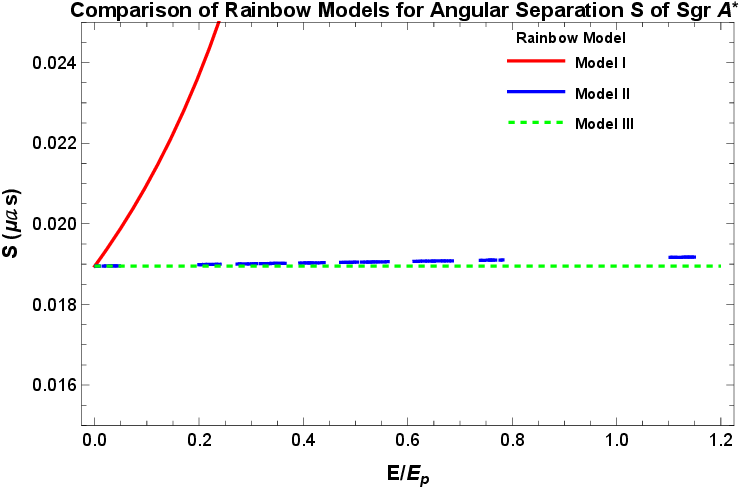}
		\end{subfigure}
  
		\caption{(a) Variation of the angular separation $S (\mu as)$ for different functional forms of the Rainbow functions, with all other parameters held fixed, in the context of M87$^*$. (b) Variation of the angular separation $S (\mu as)$ for different functional forms of the Rainbow functions, with all other parameters held fixed, in the context of Sgr A$^*$. }
		\label{fig15}
\end{figure*}

\begin{figure*}[htbp]
 \captionsetup[subfigure]{labelformat=simple}
    \renewcommand{\thesubfigure}{(\alph{subfigure})}
		
            \begin{subfigure}{.4\textwidth}
			\caption{}\label{16a}
			\includegraphics[height=3in, width=3in]{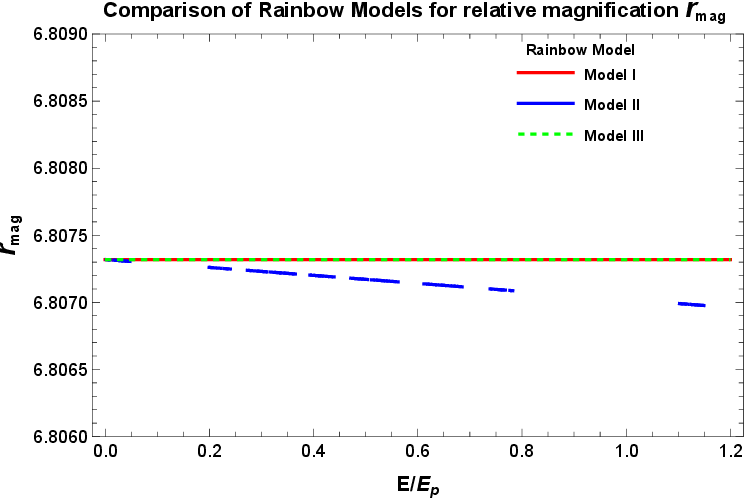}
		\end{subfigure}
  
		\caption{(a) Variation of the relative magnification $r_{mag}$ for different functional forms of the Rainbow functions, with all other parameters held fixed. }
		\label{fig16}
\end{figure*} 

\vspace{1em}
\textbf{Sensitivity Analysis and Robustness:}
To examine the robustness of our lensing predictions, we have computed few important key lensing observables such as the deflection angle, angular position, angular separation,  and relative magnification for each of the three models using a fixed set of physical parameters ($\gamma=0.001$,$\eta=1$, $\alpha=0.02$) and a fixed set of BH parameters ($Q=0.1$,$f_{R_0}=1.1$,$R_0=0.4$, $\lambda=1$).These parameter choices are found to yield viable and physically meaningful results.  The results, presented in Figs.~\ref{fig13}-\ref{fig16}, clearly demonstrate that:

\begin{table}[htbp]
\begin{center}
\begin{tabular}{|c|c|}
\hline
Key lensing observables & Comparison \\
\hline
$\alpha_D$& $Model~I \geq Model~II \geq Model~III$\\
\hline
$\theta_{\infty}$ & $Model~I \geq Model~II \geq Model~III$\\
\hline
$S$& $Model~I \geq Model~II \geq Model~III$\\
\hline
$r_{mag}$& $Model~I \geq Model~III\geq Model~II$\\
\hline
\end{tabular}
\end{center}
\caption{Comparison of the lensing observables in between Model~I , Model~II and Model~III }\label{new3}
\end{table}

\begin{itemize}
\item  Variation of the strong deflection angle $\alpha_{D}$ and strong lensing observables angular position $\theta_{\infty}$, angular separation $S$, and relative magnification $r_{\text{mag}}$ for three different functional forms of the Rainbow functions,  as Model I, Model II, and Model III. This comparison demonstrates how the predictions of gravitational lensing are influenced by the specific choice of modified dispersion relations in the framework of Rainbow Gravity.

\item  From Fig.~\ref{13a}, it is observed that strong deflection angle $\alpha_{D}$ , srong lensing observables angular position $\theta_{\infty}$, and angular separation $S$,  for the Model~I are much more than the cases of Model~II and Model~III , as shown in Table \ref{new3}, where as the  relative magnification $r_{mag}$ for the Model~I  and Model~III  almost coincide but more than the case of Model~II . This comparison highlights how gravitational lensing predictions are affected by the different functional forms of Rainbow Gravity.

    \item  The qualitative behavior of all observables remains consistent across different Rainbow function models.
    \item  However, the quantitative deviations from standard General Relativity are sensitive to the functional form of the Rainbow functions and the values of the parameters \( \eta, \alpha, \gamma \).
    \item  For a fixed value of \( E/E_p \), each model leads to a distinguishable pattern of lensing observables, which can in principle be used to constrain or distinguish between different MDRs.
\end{itemize}

This sensitivity analysis underscores the importance of the choice of Rainbow functions and justifies the need to study multiple models to establish the model-independent features of Rainbow Gravity. Our comparative results confirm that the overall physical trends are robust, though specific numerical values vary.


 \section{Concluding remarks}\label{sec7}

We have comprehensively studied the observable signatures of strong gravitational lensing by BHs in $F(R)$-EH Rainbow gravity in the context of various supermassive BHs at the centers of different galaxies. Our results demonstrate how the charge parameter $Q$ and the EH parameter $\lambda$, along with the effects of Rainbow gravity, influence key lensing parameters, including the photon sphere radius ($r_{ph}$), the critical impact parameter ($u_{ph}$), and the strong deflection angle ($\alpha_D$).  

Our study encompasses various observable quantities, including the image separation ($S$), the angular position of the innermost relativistic image ($\theta_{\infty}$), the relative magnification ($r_{\text{mag}}$), the time delays ($\Delta T_{2,1}$) between two relativistic images, and Einstein rings. We have specifically examined these strong lensing observables in the context of various supermassive BHs, including $M87^{*}$ and $Sgr A^{*}$, within $F(R)$-EH Rainbow gravity, and assessed the phenomenological differences compared to BHs in Einstein gravity. 
  Firstly, we have derived the null geodesic equations for the  BH spacetime in $F(R)$-EH Rainbow gravity using the Hamilton-Jacobi action. Subsequently, we have utilized these equations to determine the photon sphere radius ( $r_{ph}$).
It is seen that the radius of the photon sphere  $\mathit{r_{ph}}$ increases with the EH parameter $\lambda$, keeping the remaining other parameter fixed. Similarly, the radius of the photon sphere ( $r_{ph}$) decreases with the parameter $Q$, keeping other parameters fixed.
We numerically and graphically obtained the strong lensing coefficients $\mathit{\bar{a}}$, $\mathit{\bar{b}}$, and $\mathit{u_{ph}/R_s}$ for the  BH in $F(R)$-EH Rainbow gravity. It is observed that the coefficients $\mathit{\bar{a}}$, $\mathit{\bar{b}}$, and $\mathit{u_{ph}/R_s}$ increases with the EH parameter $\lambda$, keeping the remaining other parameter fixed. Similarly, these coefficients increase with the parameter $Q$, keeping other parameters fixed.
We analyzed the strong deflection angle $\alpha_D$ for BH in $F(R)$-EH Rainbow gravity using the strong lensing coefficients. It is evident that $\alpha_D$ decreases with the EH parameter $\lambda$, keeping the remaining other parameter fixed, suggesting that the EH parameter $\lambda$ enhances gravitational bending effects. Furthermore, the deflection angle $\alpha_D$ decreases with the impact parameter $u$ and diverges at the critical value of the impact parameter $u = u_{ph}$ at $r = r_{ph}$. It is also seen that the deflection angle $\alpha_D$ for the BH in $F(R)$-EH Rainbow gravity is significantly greater than the case of RN BH, standard Schwarzschild BH. This suggests that a BH in $F(R)$-EH Rainbow gravity, along with its surrounding environment, may exhibit stronger and more detectable gravitational bending effects.

The lensing observables, including the angular image position \( \theta_{\infty} \) and relative magnification \( r_{\text{mag}} \), increase with the EH parameter \( \lambda \) while keeping other parameters fixed. In contrast, \( r_{\text{mag}} \) decreases as the charge parameter \( Q \) increases. Additionally, both \( \theta_{\infty} \) and the angular separation \( S \) are slightly larger for the BH in \( F(R) \)-EH Rainbow gravity compared to standard RN and Schwarzschild BHs. Additionally, the outermost Einstein ring radius \( \theta^E_{1} \) increases with \( \lambda \) in the context of both M87* and Sgr A* supermassive BHs. A key astrophysical consequence is the time delay \( \Delta T_{2,1} \) between two relativistic images, which is significantly higher for these BHs (e.g., \( \sim 3854.19 \) minutes for NGC 7768) compared to RN (\( \sim 3350.04 \) minutes) and Schwarzschild (\( \sim 3582.72 \) minutes) BHs. These distinctions indicate that BH in \( F(R) \)-EH Rainbow gravity can be identified through strong gravitational lensing observations, distinguishing them from other astrophysical BHs. Keeping the other parameters fixed and using the observable $\theta_{\infty}$, which represents the angular radius of the BH shadow due to strong lensing, we impose constraints on the parameter $\lambda/M^2$ characterizing the $F(R)$-EH Rainbow gravity model: For M87$^*$, the allowed range is $0 \leq \lambda/M^2 \leq 6.86 \times 10^5$. For Sgr A$^*$, the constraint is $0 \leq \lambda/M^2 \leq 1.45 \times 10^5$. These parameter ranges are relatively large and depend on the choice of other parameters. However, they remain finite, indicating that the spacetime is viable according to constraints reported in existing literature.
The sensitivity of the results has been analyzed to assess how robust the lensing predictions are with respect to changes in the functional forms of the Rainbow functions. Strong gravitational lensing and its key observables have been investigated under the combined modifications of $F(R)$ gravity, Euler–Heisenberg (EH) electrodynamics, and Rainbow Gravity, which together lead to new qualitative effects not present in the individual models.

Our study indicates that BH in F(R)-EH Rainbow gravity could be more readily detectable using current strong gravitational lensing techniques. Resolving the outermost image would allow for distinguishing these BHs from their counterparts in standard RN and Schwarzschild BHs, enabling precise characterization through existing observational methods.  
Key observational signatures, including variations in Einstein ring size and time delay, provide a promising avenue for identifying and analyzing the effects of F(R)-EH Rainbow gravity. These findings underscore its potential as a viable framework for understanding the environment surrounding BH.

\section*{Acknowledgements}
SN would like to thank CSIR, Govt. of India, for providing a junior research fellowship (No. 08/0003 (17177) / 2023-EMR-I).




\bibliographystyle{apsrev4-1}%
\bibliography{rm.bib}

\end{document}